\documentclass[reprint,aps,pra,superscriptaddress,amsmath,amssymb,longbibliography]{revtex4-1}
\usepackage{graphicx}% Include figure files
\usepackage{dcolumn}% Align table columns on decimal point
\usepackage{bm}% bold math
\usepackage{braket} %Dirac notation
\usepackage{bbold}
\usepackage[colorlinks=true,urlcolor=blue,citecolor=blue,linkcolor=blue,bookmarks=false, pdfstartview={FitH}]{hyperref}
\begin{document}
\title{Surface vibrational modes of the topological insulator Bi$_2$Se$_3$ observed by Raman spectroscopy}
\author{H.-H.~Kung}
\email{skung@physics.rutgers.edu}
\affiliation{Department of Physics \& Astronomy, Rutgers University, Piscataway, NJ 08854, USA}
\author{M.~Salehi}
\affiliation{Department of Physics \& Astronomy, Rutgers University, Piscataway, NJ 08854, USA}
\affiliation{Department of Materials Science and Engineering, Rutgers University, Piscataway, NJ 08854, USA}
\author{I.~Boulares}
\affiliation{Department of Physics, University of Michigan, Ann Arbor, Michigan 48109-1040, USA}
\author{A.~F.~Kemper}
\affiliation{Department of Physics, North Carolina State University, Raleigh, North Carolina 27695, USA}
\author{N.~Koirala}
%\altaffiliation[Current address: ]{Department of Physics, Massachusetts Institute of Technology, Cambridge, Massachusetts 02139, USA}
\author{M.~Brahlek}
\affiliation{Department of Physics \& Astronomy, Rutgers University, Piscataway, NJ 08854, USA}
%\altaffiliation[Current address: ]{Department of Materials Science and Engineering, Pennsylvania State University, University Park, PA 16802, USA}
%
\author{P.~Lo\v{s}\v{t}\'{a}k}
\affiliation{Faculty of Chemical Technology, University of Pardubice, Studentska 573, 53210 Pardubice, Czech Republic}
\author{C.~Uher}
\author{R.~Merlin}
\affiliation{Department of Physics, University of Michigan, Ann Arbor, Michigan 48109-1040, USA}
\author{X.~Wang}
%\altaffiliation[Current address: ]{Department of Physics, University of Science and Technology, Beijing, China}
\author{S.-W.~Cheong}
\affiliation{Department of Physics \& Astronomy, Rutgers University, Piscataway, NJ 08854, USA}
\affiliation{Rutgers Center for Emergent Materials, Rutgers University, Piscataway, NJ 08854, USA}
\author{S.~Oh}
\affiliation{Department of Physics \& Astronomy, Rutgers University, Piscataway, NJ 08854, USA}
\author{G.~Blumberg}
\email{girsh@physics.rutgers.edu}
\affiliation{Department of Physics \& Astronomy, Rutgers University, Piscataway, NJ 08854, USA}
\affiliation{National Institute of Chemical Physics and Biophysics, 12618 Tallinn, Estonia}
%
%\date{\today}
\begin{abstract}
We present polarization resolved Raman scattering study of surface vibration modes in the topological insulator Bi$_2$Se$_3$ single crystal and thick films.
Besides the four Raman active bulk phonons, we observed four additional modes with much weaker intensity and slightly lower energy than the bulk counterparts.
Using symmetry analysis, we assigned these additional modes to out-of-plane surface phonons. 
Comparing with first principle calculations, we conclude that the appearance of these modes is due to $c$-axis lattice distortion and van der Waals gap expansion near the crystal surface.
Two of the surface modes at 60 and 173\,cm$^{-1}$ are associated with Raman active $A_{1g}$ bulk phonon modes, the other two at 136 and 158\,cm$^{-1}$ are associated with infrared active bulk phonons with $A_{2u}$ symmetry.
The latter become Raman allowed due to reduction of crystalline symmetry from $D_{3d}$ in the bulk to $C_{3v}$ on the crystal surface.
In particular, the 158\,cm$^{-1}$ surface phonon mode shows a Fano lineshape under resonant excitation, suggesting interference in the presence of electron-phonon coupling of the surface excitations.
%In addition, we observed two weak features at 67 and 126\,cm$^{-1}$ likely corresponding to in-plane surface vibrational modes.
\end{abstract}
%
%\date{\today}
\maketitle
\section{Introduction}
\begin{table*}[t]
	\caption{\label{table:sample}The list of single crystal and films measured in this study.}
	\setlength{\tabcolsep}{8pt}
	\setlength{\extrarowheight}{3pt}
	\vspace{5pt}
	\begin{tabular}{l l l l}
		\hline\hline
		Sample\,\# & Composition & Description & Growth\\ 
		\hline
		\#2 & Bi$_2$Se$_3$ & 50\,QL thick film & MBE\\
		\#8 & (Bi$_2$Se$_3$)$_m$(In$_2$Se$_3$)$_n$ & 50\,nm superlattice with (m,n)=(5,5) & MBE\\
		\#10 & (Bi$_2$Se$_3$)$_m$(In$_2$Se$_3$)$_n$ & 50\,nm superlattice with (m,n)=(10,5) & MBE\\
		\#13 & Bi$_{1.95}$In$_{0.05}$Se$_3$ & single crystal with indium doping & Bridgman\\
		\#14 & Bi$_2$Se$_3$ & pristine single crystal & Bridgman\\
		\#A & Bi$_2$Se$_3$ & pristine single crystal & Bridgman\\
		\hline\hline
	\end{tabular} 
\end{table*}

Topological insulators (TIs) are a new class of quantum matter characterized by linearly dispersed spin polarized gapless surface states within the bulk band gaps~\cite{Fu2007A,Zhang2009,Hsieh2009,Xia2009,Checkelsky2009,Bianchi2010,Beidenkopf2011,Hasan2011}, which may lead to realization of novel phenomena and applications such as spintronics and quantum computing~\cite{Fu2008,Qi2008,Qi2009,Hasan2010,Yu2010,Raghu2010,Hasan2011,Qi2011,Wang2013B,Grover2014}.

Despite the topological protection, the surface states away from the Dirac point suffer from hexagonal warping effect, resulting in increased scattering rate at the TI surface~\cite{Butch2010B,Pan2012,Valla2012}.
Among many possible inelastic scattering mechanisms, electron-phonon interaction is especially important due to its direct impact on device applications at finite temperature~\cite{Parente2013,Costache2014}.
In particular, the self-energies and symmetries of the surface vibrational modes are essential for modeling the possible relaxation channels of the surface state excitations.

Theoretical modeling of surface lattice dynamics was first developed by Lifshitz and Rosenzweig~\cite{Lifshitz1948,Lifshitz1956}, and later expanded by various workers~\cite{Wallis1957,Wallis1959,Benedek1991,Wallis1994}.
The basic idea is to consider the free surface as a perturbation of an infinite lattice, and therefore to derive the surface modes from the spectrum of bulk vibrations.
As a result, the frequencies of atomic vibration modes at the surface are modified to a smaller value than in the bulk at the Brillouin zone center ($\Gamma$ point).
If there is a gap in the phonon density-of-state (DOS) and with large enough distortion, the surface phonon DOS can be entirely separated from the bulk~\cite{Lifshitz1948,Wallis1959}.
Such modes are long lived and localized at the surface, where the dispersion can be quite different than the bulk~\cite{Lagos2017}.
However, it is often experimentally challenging to distinguish surface signal from the overwhelmingly stronger intensity contribution of the bulk.
Moreover, if the surface vibration mode is not completely gapped out from the bulk spectrum, then the surface and bulk modes are indistinguishable.
Instead, the ``bulk phonon'' acquires only a slight energy shift near the crystal surface.
Notice that the surface modes originate from abrupt termination of the lattice restoring force across bulk/vacuum interface in a semi-infinite crystal, and should not be confused with the phonons in quasi-2D ultrathin samples, which are almost decoupled from the underlying substrate of a different material~\cite{Zhao2011,Zhang2011,Chis2012,Humlicek2014,Eddrief2014}.
%Surface probes such as the inelastic helium atom scattering (HAS) and electron energy loss spectroscopy (EELS) are often used to maximize the surface to bulk phonon signal ratio~\cite{Fisher1972,Benedek1992,Toennies1991}.

Bi$_2$Se$_3$ is one of the most studied TI due to its relatively simply band structure, i.e., a single Dirac cone within the 0.3\,eV bulk band gap, much larger than the thermal energy at the room temperature.
While the bulk phonon modes have been extensively studied in Bi$_2$Se$_3$ single crystals~\cite{Kohler1974,Richter1977,LaForge2010,Zhao2011,Zhang2011,Gnezdilov2011,Kim2012,Humlicek2014,Irfan2014,Eddrief2014,Yan2015,Zhang2016}, only a few papers have reported studies of the surface vibration modes.
Zhu and coworkers observed strong Kohn anomaly at about $2k_F$ using helium atom scattering (HAS)~\cite{Zhu2011}, and deduced the interaction between surface phonon and the Dirac electrons to be much stronger than the values previously reported by angle-resolved photoemission spectroscopy (ARPES) measurements~\cite{Zhu2012,Howard2013,Hatch2011,Pan2012}, suggesting that the electron-phonon coupling on TI surface may be more complex than anticipated.
Time-resolved ARPES study of single crystals reported the observation of one $A_{1g}$ bulk phonon at about 74\,cm$^{-1}$, and an additional mode with slightly lower energy consistent with what was suggested by transport measurements~\cite{Costache2014}.
This mode was interpreted as a surface phonon associated with the observed $A_{1g}$ bulk phonon~\cite{Sobota2014}.
However, alternative results have also been reported~\cite{Hatch2011,Pan2012,Chen2013,Kondo2013}, suggesting the existence of multiple phononic decaying channels which may depend on details of sample preparation.
Electron energy loss spectroscopy (EELS) study has distinguished a weak mode at about 160\,cm$^{-1}$ in Bi$_2$Se$_3$, which was assigned to the surface vibration mode associated with an $A_{1g}$ bulk phonon~\cite{Kogar2015}.
The Raman scattering work on bulk single crystal~\cite{Gnezdilov2011} and exfoliated nano-crystals reported several additional features, and were attributed to infrared active phonon modes becoming Raman active due to inversion symmetry breaking at crystal surface~\cite{Eddrief2014,Zhao2011}.
%On the other hand, another broad feature at about 245\,cm$^{-1}$ (Fig.~\ref{fig:1}(b), marked by arrow) was observed in MBE grown thick films, and was assigned to the 2D stretching mode of Se atoms on the surface~\cite{Glinka2015A}.

To date, different surface modes were measured by several distinct spectroscopies, with slight discrepancies between the results and interpretations.
To resolve the discrepancy, it is desirable to study all surface vibration modes within one technique that provides both high energy resolution and symmetry information.

\begin{figure*}[t]
	\includegraphics[width=17cm]{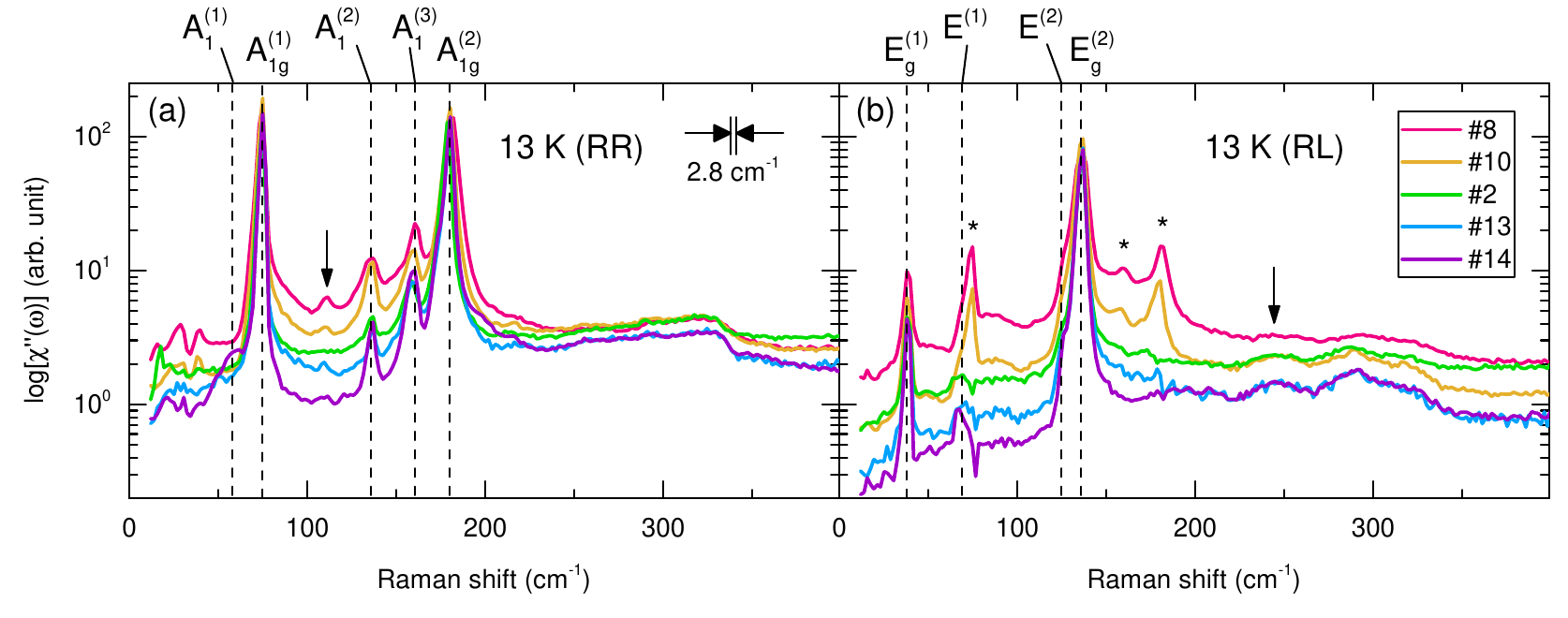}
	\caption{\label{fig:1}(Color online)
		The Raman response function $\chi^{\prime\prime}(\omega)$ measured in the (a)~RR and (b)~RL scattering geometry at 13\,K with 532\,nm excitation from various Bi$_2$Se$_3$ samples as described in Table~\ref{table:sample}, plot in semi-log scale.
		The dashed lines label the observed phonon modes as tabulated in Table~\ref{table:mode}.
		(a)~The mode at 110\,cm$^{-1}$ indicated by arrow is due to the phonon signal from $\alpha$-In$_2$Se$_3$ layers~\cite{Lewandowska2001}.
		The asterisks mark the phonon modes with $A_{1g}$ and $A_1$ symmetries, appear in RL geometry due to indium atom diffusion.
		The instrumental resolution of 2.8\,cm$^{-1}$ is shown.
	}
\end{figure*}

Raman spectroscopy is a conventional tool for studying surface phonon modes~\cite{Esser1999,Liebhaber2014}.
%The advantages of Raman spectroscopy is the wider accessible energy range compared to HAS, much higher energy resolution compared to EELS, and the additional information on the symmetries of the phonon modes.
Here, we use high resolution polarization resolved Raman spectroscopy to study the vibrational modes in Bi$_2$Se$_3$ samples.
We focus our study to the bulk single crystals, which are unexposed to air or any chemicals.
In addition to the four Raman active bulk phonons, we observed 6 additional modes with about 20--100 times weaker intensities compared to the bulk phonons [Fig.~\ref{fig:1}].
By comparing the data to the results obtained by the complementary spectroscopic techniques and the calculations, we assign the observed additional modes to surface phonons arising from out-of-plane lattice distortion near the crystal/film interface.

This paper is organized as follows. 
In Sec.~\ref{sec:Exp}, we introduce the experiments including sample preparations and the Raman probe.
In Sec.~\ref{sec:Results}, we present the low temperature polarized Raman spectra of bulk and thin film Bi$_2$Se$_3$ samples.
Sec.~\ref{sec:Discussion} discusses the symmetries and microscopic views of the surface vibration modes.
Finally, we conclude our discussions in Sec.~\ref{sec:Conclusion}.
Details of data analysis are given in Appendix.
\section{Experimental setup}\label{sec:Exp}
Table~\ref{table:sample} lists 6 Bi$_2$Se$_3$ single crystals and films measured in this study.
The bulk single crystals were grown by modified Bridgman method~\cite{Lostak1990,Dai2016}.
The thin film samples were epitaxially grown on Al$_2$O$_3$ (0001) substrates in a custom designed molecular beam epitaxy (MBE) chamber~\cite{Brahlek2012,Bansal2012}.
They were immediately transfered into a cryostat after taking out of MBE chamber.

The superlattice thin films of (Bi$_2$Se$_3$)$_m$(In$_2$Se$_3$)$_n$ are grown along (0001) surface~\cite{Brahlek2012}, where each primitive cell consists of $m$\, quintuple layer (QL) Bi$_2$Se$_3$ and $n$\,QL In$_2$Se$_3$, with each QL being $\approx 1$\,nm thick.
Notice that the light penetration depth in Bi$_2$Se$_3$ within energy range of current study is about 10\,nm~\cite{McIver2012}.
Therefore, the signal is dominated by scattering from the first few QLs of Bi$_2$Se$_3$, and the scattering volume in the superlattice samples is practically the same as bulk.

Bi$_2$Se$_3$ has a rhombohedral crystal structure with the $D_{3d}$ point group symmetry.
The irreducible representations and Raman selection rules are given in Table~\ref{table:IRR}.
With five atoms in a primitive unit cell, there are a total of three acoustic and 12 optical bulk phonon branches.
At the $\Gamma$-point, the irreducible representations of the Raman active phonons are $2A_{1g}+2E_g$, and the infrared active phonons are $2A_{2u}+2E_u$~\cite{Kohler1974,Richter1977}.
These bulk phonon modes have been measured by Raman and infrared spectroscopies~\cite{Kohler1974,Richter1977,LaForge2010,Zhao2011,Zhang2011,Gnezdilov2011,Kim2012,Humlicek2014,Irfan2014,Eddrief2014,Yan2015,Zhang2016}, and the values reported in Ref.~\cite{LaForge2010} and Ref.~\cite{Gnezdilov2011} are listed in Table~\ref{table:mode}.

The crystal naturally cleaves along the (111) surface terminated at Se atoms, forming optically flat QLs weakly bonded by van der Waals force~\cite{Kohler1974}.
The surface QL has the symmorphic $P6mm$ wallpaper group symmetry (two dimensional crystallographic point group $C_{6v}$)~\cite{Terzibaschian1986,Li2013Mar,Slager2013}.
Since the surface layer phonon modes in Bi$_2$Se$_3$ are not perfectly localized and decay into the bulk, it is more appropriate to analyze our experimental results within the layer group $P3m1$ (crystallographic point group $C_{3v}$, which is a subgroup containing common symmetry operators of $D_{3d}$ and $C_{6v}$ groups)~\cite{Li2013Mar}.

All Raman scattering measurements are taken from \textit{ab} surfaces freshly cleaved or grown immediately prior to the measurements.
Sample\,\#2--14 are measured in a quasi-backscattering geometry in a continuous He-flow optical cryostat.
A glove bag with controlled dry nitrogen gas environment was sealed to the cryostat loading port. 
After purging the bag to the desired conditions, the single crystals were cleaved in the glove bag immediately before loading into the cryostat for cooling, without exposure to air. 
We use $\lambda_L=$532\,nm solid state laser for excitation, where the spot size is roughly $50\,\mu m$.
The scattered light was analyzed and collected by a custom triple-grating spectrometer equipped with a liquid nitrogen cooled CCD detector.

As for the data collected from sample\,\#A, measurements were done in a back-scattering geometry from a cold-finger cryostat.
An argon ion laser and a Ti:Sapphire laser were used as sources, where the spot sizes are roughly 35 and 55\,$\mu m$, respectively. 
The scattered light was collected using a triple stage spectrometer (Dilor XY) and imaged on a CCD camera.

All spectra shown were corrected for the spectral response of the spectrometer 
and CCD to obtain the Raman intensity $I_{\mu\nu}(\omega,T)$, which is related to the Raman response function $\chi_{\mu\nu}^{\prime\prime}(\omega,T)$ by the Bose factor $n(\omega,T)$:
$I_{\mu\nu}(\omega,T)=[1+n(\omega,T)]\chi_{\mu\nu}^{\prime\prime}(\omega,T)$.
Here, $\mu$ ($\nu$) denotes the polarization of incident (scattered) photon, $\omega$ is energy and $T$ is temperature.
The scattering geometries used in this experiment are denoted as $\mu\nu = RR, RL, XX$ and $YX$, which is short for $\overline{z}(\mu\nu)z$ in Porto's notation.
$\text{R}=\text{X}+i\text{Y}$ and $\text{L}=\text{X}-i\text{Y}$ denotes the right- and left-circular polarizations, respectively, where X (Y) denotes linear polarization parallel (perpendicular) to the plane of incidence.
The irreducible representations of the $D_{3d}$ and $C_{3v}$ groups corresponding to these scattering geometries are listed in Table~\ref{table:IRR}.
Notice that in both the $D_{3d}$ and $C_{3v}$ groups, the phonon intensities do not depend on the orientation of the crystallographic axis.
The notations $X$ and $Y$ have no reference to the crystallographic \textit{a} and \textit{b} axes.
In order to avoid confusion with the weak surface modes, possible polarization leakage arising from optical elements are removed from presented data with a procedure described in Appendix.

\begin{table}[b]
	\caption{\label{table:IRR}
		The Raman selection rules in the bulk and on the surface of Bi$_2$Se$_3$~\cite{Ovander1960,Cardona1982}.
		Upon the reduction of symmetry from point group $D_{3d}$ to $C_{3v}$, the $A_{1g}$ and $A_{2u}$ irreducible representations merge into $A_1$, $A_{2g}$ and $A_{1u}$ merge into $A_{2}$, $E_{g}$ and $E_{u}$ merge into $E$.~\cite{Koster1963}}
	\setlength{\tabcolsep}{8pt}
	\setlength{\extrarowheight}{3pt}
	\vspace{5pt}
	\begin{tabular}{l l l}
		\hline\hline
		Scattering & Bulk & Surface \\
		geometry & ($D_{3d}$) & ($C_{3v}$) \\
		\hline
		RR & $A_{1g}+A_{2g}$ & $A_{1}+A_{2}$ \\ 
		RL & $2E_g$ & $2E$ \\
		XX & $A_{1g}+E_g$ & $A_{1}+E$ \\
		YX & $A_{2g}+E_g$ & $A_{2}+E$ \\
		\hline\hline
	\end{tabular} 
\end{table}
\section{Results}\label{sec:Results}
Figure~1 shows the Raman response function $\chi^{\prime\prime}(\omega)$, taken at 13\,K with 532\,nm excitation, plot in semi-log scale.
In order to confirm the tiny features of surface modes, we compared the results from bulk crystals and MBE films.
Figures~1(a) and 1(b) are measured with the RR and RL scattering geometries, respectively (Table~\ref{table:IRR}).
%The polarization leakage from optical elements are removed.
The dashed lines label the observed phonons as tabulated in Table~\ref{table:mode}.
The strong modes at 72 and 174\,cm$^{-1}$ in RR scattering geometry are the bulk $A_{1g}$ phonons of Bi$_2$Se$_3$ [Fig.~1(a)], and the strong modes centered at 37 and 132\,cm$^{-1}$ in RL are the bulk $E_g$ phonons [Fig.~1(b)], consistent with previous Raman studies~\cite{Zhang2011,Gnezdilov2011} and calculations~\cite{Wang2012}.

\begin{figure}[t]
	\includegraphics[width=8cm]{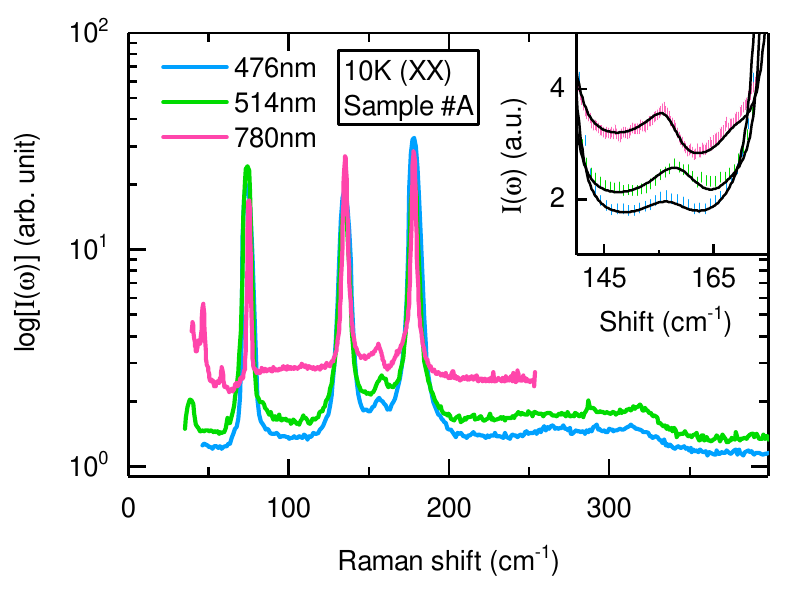}
	\caption{\label{fig:2}(Color online)
		The signal intensity in the XX scattering geometry, measured at 10\,K from a bulk Bi$_2$Se$_3$ single crystal, plot in semi-log scale.
		The blue, green and pink lines corresponds to laser excitation energy of 476, 514 and 780\,nm, respectively.
		Inset: enlarged plot around the $A_1^{(3)}$ mode. The black lines are fit to Fano lineshape (Eq.~\ref{eq:Fano}).
	}
\end{figure}

The broad feature at about 330\,cm$^{-1}$ in RR is possibly due to second-order scattering of the $A_{1g}^{(2)}$ phonon, broadened due to the large downward dispersion of the phonon branch~\cite{Wang2012}.
Similarly, the broad feature observed around 300\,cm$^{-1}$ in RL is assigned to two-phonon excitation, $A_{1g}^{(2)}+E_g^{(2)}$.
The broad feature at about 245\,cm$^{-1}$ [Fig.~\ref{fig:1}(b), marked by arrow] was previously assigned to the 2D stretching mode of Se atoms on the surface~\cite{Glinka2015A}.
However, we do not observe the reported resonance effect of this mode with near-infrared excitation [Fig.~\ref{fig:2}].
Notice that this mode energy is also consistent with the two-phonon excitation of $A_{1g}^{(2)}+E_g^{(1)}$.

In order to distinguish the broad features from electronic origin, such as excitations from the topological surface states, we compared the results with indium doped Bi$_2$Se$_3$ in Fig.~\ref{fig:1}.
Indium doping was shown to increase the carrier density and suppress the topological surface states in Bi$_2$Se$_3$~\cite{Brahlek2012,Wu2013}.
Here, we collected data from bulk single crystals and MBE grown In$_2$Se$_3$/Bi$_2$Se$_3$ superlattices, where indium doping is achieved through diffusion in the superlattices~\cite{Lee2014}.
In all indium doped samples, the broad features show the same intensity, suggesting their origin unrelated to the topological surface states.
This feature is slightly weaker in the superlattice sample\,\#8, despite the first-order phonon modes are still sharp and strong.
However, this is likely mainly due to the indium atom diffusion into the Bi$_2$Se$_3$ layer, which breaks the translation symmetry, and therefore further broadens the multi-phonon mode.
%The indium atom diffusion also causes the non-negligible intensity of $A_{1g}$ and $A_1$ symmetry modes present in RL for both superlattice samples (Fig.~\ref{fig:1}(b), marked by asterisks).
The diffused indium atoms also lower the local crystal symmetry in the Bi$_2$Se$_3$ layers, and therefore allows vibration modes with $A_{1g}$ and $A_1$ symmetries to appear in the RL geometry, which is otherwise forbidden for the crystal symmetry of Bi$_2$Se$_3$ [Fig.~\ref{fig:1}(b), marked by asterisks].
The small feature at 110\,cm$^{-1}$ in RR is due to a strong phonon of $\alpha$-In$_2$Se$_3$ layers~\cite{Lewandowska2001} (indicated by arrow in Fig.~1(a)).

In addition to the strong bulk first-order Raman phonons and the broad features, we see additional sharp modes that are about 20 times weaker than the bulk phonons.
In Fig.~\ref{fig:1}(a), two such features at 136 and 158\,cm$^{-1}$ are seen in all samples in RR scattering geometry, labeled $A_1^{(2)}$ and $A_1^{(3)}$, respectively.
In the bulk single crystal sample \#14, we observed a mode at about 60\,cm$^{-1}$, which we label as $A_1^{(1)}$.
We associate these three features with vibration modes at the crystal surface, to be discussed in the RR polarization for the Sample\,\#14 in the next section.
We also noticed several sharp features below 50\,cm$^{-1}$ in sample\,\#8 and \#10 in RR, which are possibly zone folded phonons.
To confirm this requires further studies, and is beyond the scope of this paper.
In the RL scattering geometry, we observed two weak features at 67 and 126\,cm$^{-1}$, labeled $E^{(1)}$ and $E^{(2)}$, respectively [Fig.~\ref{fig:1}(b)].
The energy of these modes are close to the strong bulk phonons, and therefore require higher resolution to distinguish them.

In Fig.~\ref{fig:2} are the Raman spectra of the bulk sample at different excitation wavelengths at 10\,K. 
The spectra were obtained in the XX polarization. 
As in Fig.~\ref{fig:1}, we observe an additional peak at 158\,cm$^{-1}$ which we refer to as $A_1^{(3)}$. 
However, note that the mode is more asymmetric when 780 nm excitation wavelength is used.
This is an indication that the $A_1^{(3)}$ phonon is interacting with a continuum.
\begin{figure}[t]
	\includegraphics[width=8cm]{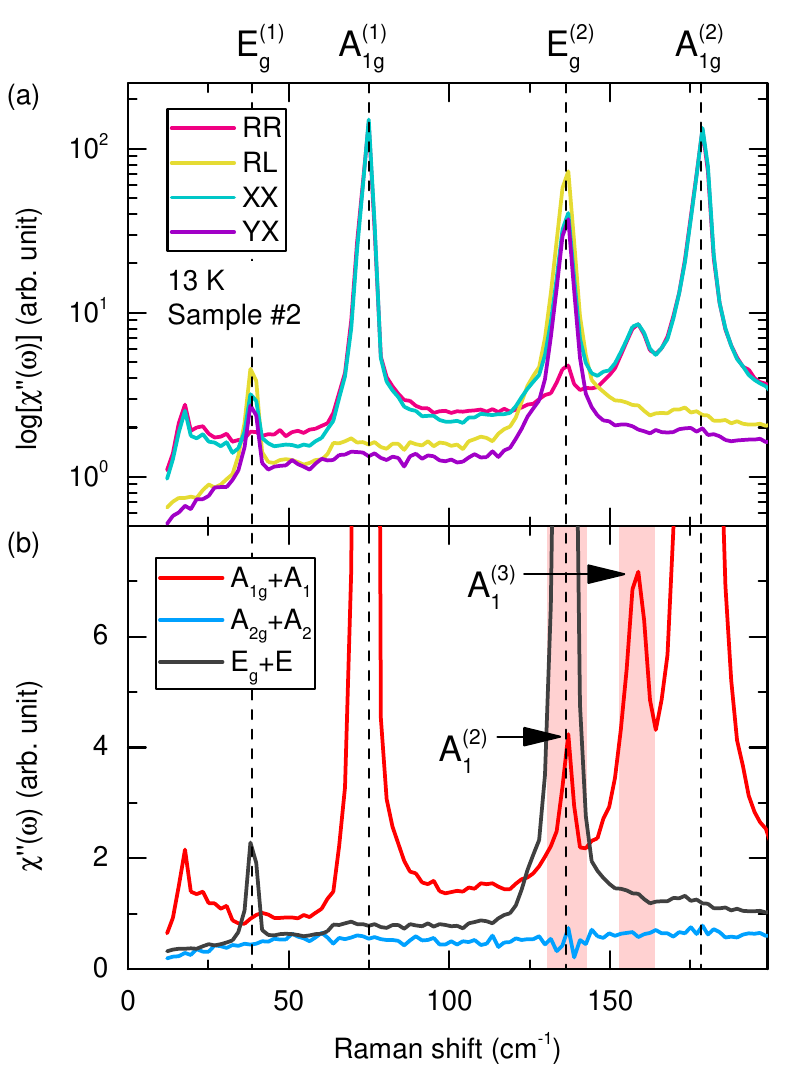}
	\caption{\label{fig:sym}(Color online)
		(a)~The Raman spectra taken in all four scattering geometries at 13\,K with 532\,nm excitation from a Bi$_2$Se$_3$ thick film, plotted on a semi-log scale.
		(b)~The Raman response of different symmetry channels, obtained from data in (a).
		The bulk phonons are marked by dashed lines, whereas the surface modes are indicated by arrows and shaded in red.
	}
\end{figure}

To further understand the observed phonon modes, we measure the Raman response in four scattering geometries of the $D_{3d}$ and $C_{3v}$ point group as listed in Table~\ref{table:IRR} [Fig.~\ref{fig:sym}(a)].
The intensity contributed by each symmetry channel in different scattering geometries are dictated by the Raman tensors~\cite{Cardona1982,Ovander1960} and the results for $D_{3d}$ and $C_{3v}$ groups are listed in Table~\ref{table:IRR}.
Therefore, by obtaining polarized Raman spectra in four proper scattering geometries, we can separate the measured Raman response from each symmetry channel.
\begin{eqnarray}
\chi^{\prime\prime}_{A1g}(\omega)+\chi^{\prime\prime}_{A1}(\omega)=&&\chi^{\prime\prime}_{\text{XX}}(\omega)-\frac{1}{2}\chi^{\prime\prime}_{\text{RL}}(\omega) \nonumber\\
\chi^{\prime\prime}_{A2g}(\omega)+\chi^{\prime\prime}_{A2}(\omega)=&&\chi^{\prime\prime}_{\text{YX}}(\omega)-\frac{1}{2}\chi^{\prime\prime}_{\text{RL}}(\omega) \\
\chi^{\prime\prime}_{Eg}(\omega)+\chi^{\prime\prime}_{E}(\omega)=&&\frac{1}{2}\chi^{\prime\prime}_{\text{RL}}(\omega) \nonumber
\end{eqnarray}

The results are shown in Fig.~\ref{fig:sym}(b).
%The response from $E_g$ and $E$ channels are cleanly obtained from RL scattering geometry.
%{Koningstein1968,Ko2010,Liu1993,Rho2010,Kung2015,Riccardi2016}.
We notice that no lattice vibrational mode is observed in the $A_{2g}$ and $A_{2}$ symmetry channels.
This is because the Raman tensors for these two channels are antisymmetric and commonly correspond to pseudo-vector-like excitations~\cite{Ovander1960,Shastry1991,Khveshchenko1994}, which is forbidden for phononic Raman scattering in Bi$_2$Se$_3$.
Since the signal in $A_{2g}$ and $A_{2}$ channels are expected to be zero, we can claim that all vibration modes appearing in RR have either $A_{1g}$ or $A_{1}$ symmetry (Table~\ref{table:IRR}).
%justifying our mode assignments in Table~\ref{table:mode}
%Notice that contributions from the bulk ($A_{1g}$ and $A_{2g}$) and surface ($A_1$ and $A_2$) cannot be disentangled simply by symmetry analysis.

The $A_1^{(2)}$ mode happens to have energy very close to the $E_g^{(2)}$ phonon, making it particularly difficult for spectroscopic experiments to distinguish.
Here, we utilize the symmetry properties to separately detect them with polarized light.
The polarization leakage of optical elements are precisely measured and removed (Appendix), and thereby excluding the possibility of $A_1^{(2)}$ being a trivial polarization leakage from the $E_g^{(2)}$ phonon.

\begin{figure}[t]
	\includegraphics[width=8cm]{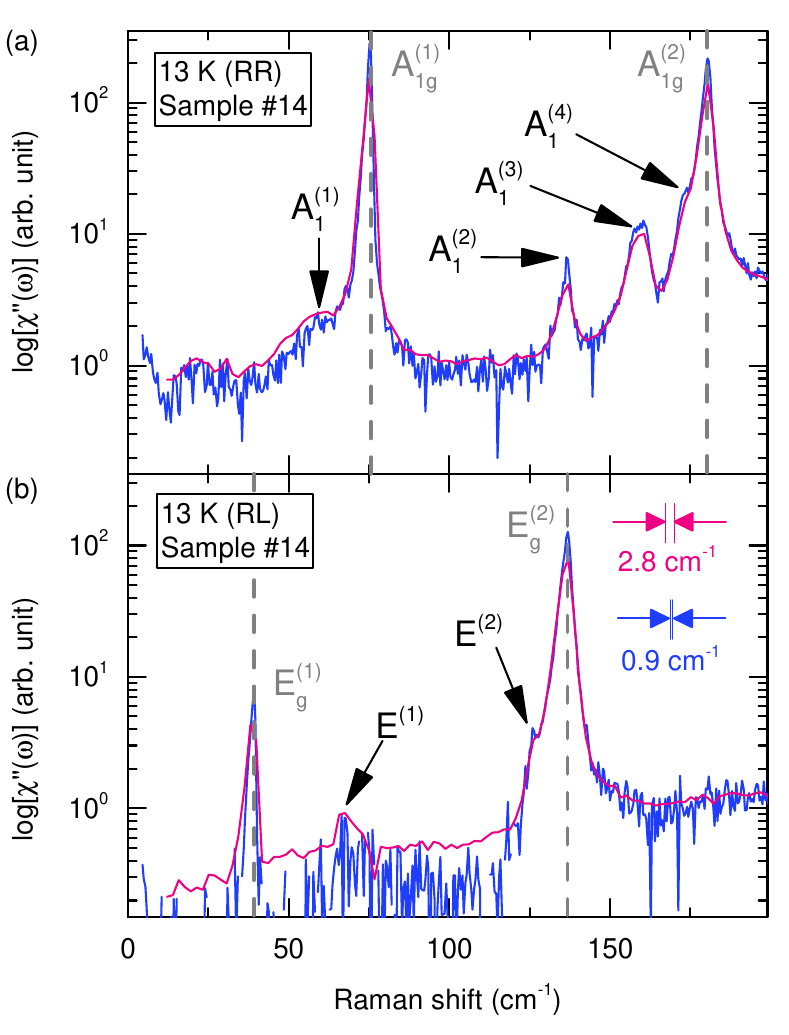}
	\caption{\label{fig:hiRes}(Color online)
		The Raman spectra taken in the (a)~RR and (b)~RL scattering geometry at 13\,K with 532\,nm excitation from a bulk Bi$_2$Se$_3$ single crystal are plotted on a semi-log scale.
		The red and blue curves correspond to instrumental resolution of 2.8 and 0.9\,cm$^{-1}$ (as shown in (b)), respectively.
		The bulk phonons are marked by gray dashed lines.
	}
\end{figure}

To distinguish surface modes that are particularly weak and close in energy to the bulk phonons, we take high resolution spectra from a carefully prepared bulk crystal \#14, cleaved in nitrogen environment.
We show in Fig.~\ref{fig:hiRes} the spectra taken at 13\,K in RR and RL scattering geometries, where the smoother low resolution (2.8\,cm$^{-1}$) data is overlapped with the high resolution (0.9\,cm$^{-1}$) spectra.
Besides the more pronounced $A_1^{(2)}$ and $A_1^{(3)}$ modes already visible in Fig.~\ref{fig:sym}, we see a few additional features in the high resolution data:
(1)~A mode centered at 173\,cm$^{-1}$ appearing as a shoulder to the $A_{1g}^{(2)}$ bulk phonon in RR geometry [Fig.~\ref{fig:hiRes}(a)], which we designate as $A_{1}^{(4)}$.
(2)~Another mode centered at 126\,cm$^{-1}$ appearing as a shoulder to the $E_{g}^{(2)}$ bulk phonon in RL geometry [Fig.~\ref{fig:hiRes}(b)], which we designate as $E^{(2)}$.
(3)~The mode $A_{1}^{(3)}$ shows broadened peak structure. This cannot be due to splitting of an $A$-symmetry phonon, e.g., lowering of symmetry, since $A_1$ is a one-dimensional representation. 
This can be explained as due to Fano interference, which become more pronounced with infrared excitation [Fig.~\ref{fig:2}].

\section{Discussion}\label{sec:Discussion}
\begin{table*}[th]
	\caption{\label{table:mode}
		The summary of the bulk and surface phonon mode energies. 
		This work’s data is compared to the spectroscopic studies reported in Ref.~\cite{Gnezdilov2011,LaForge2010,Sobota2014,Kogar2015,Zhang2011,Kim2012,Eddrief2014,Humlicek2014,Irfan2014,Yan2015,Zhang2016,Zhao2011}, and the calculations reported in Ref.~\cite{Chen2011,Wang2012}. 
		All values are given in units of cm$^{-1}$.}
	\vspace{5pt}
	\setlength{\tabcolsep}{8pt}
	\setlength{\extrarowheight}{3pt}
	\begin{tabular}{l c c c c c}
		\hline\hline
		{} & \multicolumn{2}{c}{Experiment} & {    } & \multicolumn{2}{c}{Calculation} \\
		\cline{2-3} \cline{5-6}
		Symmetry & \multicolumn{1}{c}{This work} & \multicolumn{1}{c}{Literature} & & \multicolumn{1}{c}{LDA+SOI~\cite{Wang2012}} & \multicolumn{1}{c}{GGA+SOI~\cite{Chen2011}} \\
		\hline 
		$A_{1g}^{(1)}$ & 75 & 73~\cite{Gnezdilov2011,Zhang2011,Irfan2014,Zhang2016} &  & 77 & 64 \\		
		$A_{1g}^{(2)}$ & 180 & 175~\cite{Gnezdilov2011,Zhang2011,Irfan2014,Zhang2016} &  & 176 & 167 \\
		$E_{g}^{(1)}$ & 39 & 39~\cite{Gnezdilov2011,Zhang2011,Irfan2014,Zhang2016} &  & 41 & 39 \\
		$E_{g}^{(2)}$ & 137 & 133~\cite{Gnezdilov2011,Zhang2011,Irfan2014,Zhang2016} &  & 139 & 124 \\
		$A_{2u}^{(1)}$ & -- & N/A &  & 139 & 137 \\
		$A_{2u}^{(2)}$ & -- & N/A &  & 161 & 156 \\
		$E_{u}^{(1)}$ & -- & 61~\cite{LaForge2010} &  & 80 & 65 \\
		$E_{u}^{(2)}$ & -- & 133~\cite{LaForge2010} &  & 131 & 127 \\
		\hline
		$A_{1}^{(1)}$ & 60 & 68~\cite{Sobota2014} &  & N/A & N/A \\
		$A_{1}^{(2)}$ & 136 & 129~\cite{Gnezdilov2011} &  & N/A & N/A \\
		$A_{1}^{(3)}$ & 158 & 160~\cite{Gnezdilov2011,Kogar2015} &  & N/A & N/A \\
		$A_{1}^{(4)}$ & 173 & N/A &  & N/A & N/A \\
		$E^{(1)}$ & 67 & 68~\cite{Gnezdilov2011} &  & N/A & N/A \\
		$E^{(2)}$ & 126 & 125~\cite{Gnezdilov2011} &  & N/A & N/A \\
		\hline\hline
	\end{tabular}
\end{table*}

At the crystal surface of Bi$_2$Se$_3$, the lattice structure is distorted along \textit{c}-axis due to the abrupt reduction of the interlayer van der Waals force that binds the crystal together, and is calculated by density functional theory (DFT) to be about 10\% along $c$-axis~\cite{Sobota2014}.
%The atoms do not like to remain in the bulk coordinates, and as a result shift around.
Additionally, the observation of two-dimensional electron gas formed on Bi$_2$Se$_3$ surface also supports the picture of subsurface van der Waals gap expansion~\cite{Bianchi2010,Bianchi2012,Menshchikova2011}.
However, finite phonon DOS exist across the entire energy range in Bi$_2$Se$_3$~\cite{Wang2012}, allowing the surface modes to decay into bulk phonon modes.
Therefore, the surface mode is not entirely ``peeled off'' from the bulk.
Instead, one would expect a ``surface resonance'' with slightly lower energy than the bulk phonon.

Due to inversion symmetry breaking at the crystal interface, the surface resonance from the Raman active $A_{1g}$ and IR active $A_{2u}$ phonons are both expected to appear in the $A_1$ symmetry ($C_{3v}$ group), corresponding to out-of-plane atomic motion.
The energies of such surface modes are usually slightly lower than the corresponding bulk phonons.
This is consistent with the four $A_1$ modes we observed [Fig.~\ref{fig:hiRes}(a)].
From the energies of these $A_1$ modes, we conclude that $A_{1}^{(1)}$ and $A_{1}^{(4)}$ are associated with the bulk phonon modes $A_{1g}^{(1)}$ and $A_{1g}^{(2)}$, respectively.
The measured energy of the $A_{1}^{(1)}$ mode is somewhat different than the previously reported value of 68\,cm$^{-1}$ by time resolved ARPES~\cite{Sobota2014}, but close to what was suggested by transport measurements~\cite{Costache2014}.
We believe this difference may be partly due to surface quality variation. 
ARPES measured sample is usually cleaved in ultra high vacuum, whereas the surface in this study is cleaved in nitrogen environment.
This may also explain why this mode was not observed in the MBE samples [Fig.~\ref{fig:1}], where the sample is unavoidably exposed to air for a few minutes during the transfer between MBE chamber and Raman cryostat.
The $A_{1}^{(4)}$ mode appears as a shoulder to the $A_{1g}^{(2)}$ bulk phonon, requiring higher resolution to distinguish from the bulk mode, and therefore was overlooked in the previous Raman study~\cite{Gnezdilov2011}.

In comparison, the surface modes $A_{1}^{(2)}$ and $A_{1}^{(3)}$ have higher intensity and are better resolved.
One possibility for this difference is that the bulk counterpart of these modes are the IR active $A_{2u}^{(1)}$ and $A_{2u}^{(2)}$ phonons, as the measured energy is close to the calculated values (Table~\ref{table:mode}).
Since these bulk modes are not Raman active, we were able to better resolve the surface resonance.
Another possibility is that the phonon DOS is practically zero at these energies in the $A_1$ symmetry channel, and the surface vibration modes are truly localized.
Distinguishing these two scenarios is in fact experimentally non-trivial, especially since the experimental values of the $A_{2u}^{(1)}$ and $A_{2u}^{(2)}$ bulk phonon energies are yet unknown.

Nevertheless, both possibilities point to the surface origin of these two modes, which provide us with information on the electron-phonon coupling at the TI surface.
While the bulk phonons show little resonance effect, the $A_{1}^{(3)}$ phonon displays antisymmetric lineshape with 780\,nm excitation, reminiscent of a Fano lineshape~\cite{Fano1961} [Fig.~\ref{fig:2}, inset].
This was overlooked in previous Raman studies, and may be related to the 20\,meV ``kink'' in the topological surface state's energy dispersion curve reported by some ARPES measurements~\cite{Kondo2013,Chen2013}.
%Notice that EELS study also observed the $A_{1}^{(3)}$ surface mode at 20\,meV and a strong surface plasmon at about 25\,meV in crystals with low carrier concentrations~\cite{Kogar2015}.
%It is possible that one of the two peaks we see is due to the surface plasmon, and the energies are too close for EELS resolution to distinguish.
The observation of Fano lineshape is a clear evidence for the existence of underlying electronic continuum in the $A_1$ symmetry channel, which interacts with the $A_{1}^{(3)}$ phonon~\cite{Fano1961,Klein1983}.
The excitation dependence also suggests resonance enhancement of the electronic continuum with near-infrared wavelength, consistent with the reported surface states at about 1.6\,eV above the Fermi energy~\cite{Sobota2013,Niesner2012}.
Fitting the 780\,nm data with Eq.~4.48 in Ref.~\cite{Klein1983}:
\begin{equation}\label{eq:Fano}
I(\omega)=\frac{\pi\rho T_e^2(\omega_0-\omega-V T_p/T_e)^2}{(\omega_0-\omega+V^2 R)^2+(\pi V^2 \rho)^2}\,,
\end{equation}
yields electron-phonon interaction strength $V\approx 2.6$\,cm$^{-1}$, and phonon energy $\omega_0\approx 158$\,cm$^{-1}$.
Here we assumed the electron DOS $\rho$ is a constant in the relevant energy window, and neglect the real part of the electronic Green's function $R$.
$T_p$ and $T_e$ are the phonon and electronic continuum Raman transition matrix elements, respectively.
%However, group symmetry analysis shows that surface phonons with $A_1$ symmetry at $\Gamma$ point can scatter with an electronic continuum, if and only if the continuum is resulting from surface-to-surface excitations~\cite{Li2013Mar}.
%Therefore, the Fano lineshape of the $A_{1}^{(3)}$ phonon must be due to interaction with the topological surface states, consistent with the ARPES measurements~\cite{Kondo2013,Chen2013}.

Since the in-plane symmetries are mainly preserved as the DFT calculated atomic surface distortion is purely out-of-plane~\cite{Sobota2014}, one would not expect surface phonon with $E$ symmetry ($C_{3v}$ group) for Bi$_2$Se$_3$.
However, the in-plane bonding potential is also modified by having distortion along \textit{c}-axis, and therefore the phonon frequency at surface is still slightly different than the bulk value.
If the modification is tiny, the $E$ modes are expected to be weak and close to the bulk phonons.
In Fig.~\ref{fig:1}(b) and Fig.~\ref{fig:hiRes}(b), we can see hints of two additional modes, labeled by $E^{(1)}$ and $E^{(2)}$.
The energies of these modes are in fact close to the measured values of $E_{u}^{(1)}$ and $E_{u}^{(2)}$ bulk phonons~\cite{Richter1977,LaForge2010}, and are consistent with the previous Raman study~\cite{Gnezdilov2011} (Table~\ref{table:mode}).
However, the frequency of $E_1$ is slightly higher than $E_{u}^{(1)}$, which is against the expectation from a surface resonance.
This may reflect the fact that this is an in-plane mode, orthogonal to the lattice distortion direction.
Or, this may be indicative of non-trivial electron-phonon interaction with the surface states, and worth further studying. 

\section{Conclusion}\label{sec:Conclusion}
In conclusion, we have done systematic symmetry analysis on the temperature and excitation dependent Raman spectra from high quality, freshly cleaved or grown \textit{ab} surfaces of Bi$_2$Se$_3$ single crystal and films.
We observed in total four out-of-plane, and possibly 2 in-plane surface vibrational modes, where we tabulate the energies and symmetries in Table~\ref{table:mode}.

In particular, we reproduced the $A_1^{(1)}$ mode, which was previously observed in time resolved ARPES measurements~\cite{Sobota2014}.
The $A_1^{(1)}$ mode is interesting because it was found to couple strongly with the topological surface states, and therefore provides the main phononic decay channel for the Dirac fermions in Bi$_2$Se$_3$.
Our report of energies and symmetries of the $A_1^{(1)}$ and other surface modes affirms the validity of the surface lattice distortion model employed in Ref.~\cite{Sobota2014}.
The consistently much larger intensity for the out-of-plane vibration modes compared to in-plane modes strongly suggest that the surface lattice distortion and van der Waals gap expansion in Bi$_2$Se$_3$ is only along $c$-axis.

Lastly, the $A_{1}^{(2)}$ and $A_{1}^{(3)}$ modes have much stronger intensities compared to the other surface vibration modes, and may be candidates for localized surface phonons.
In particular, we noticed the $A_{1}^{(3)}$ mode possesses a Fano lineshape in low doped Bi$_2$Se$_3$ single crystals.
The Fano lineshape is indicative of electron-phonon coupling with the underlying electronic continuum of the same symmetry, important for understanding the relaxation and scattering of surface state excitations.
Here, we found a resonance effect to the Fano lineshape with 780\,nm excitation, suggesting the onset of the electronic continuum in $A_1$ symmetry has excitation dependence.
This explains the inconsistent surface electron-phonon coupling constant found in previous ARPES studies~\cite{Pan2012,Hatch2011}.
The excitation dependence also confirms the existence of unoccupied surface states at about 1.6\,eV above the Fermi energy, which enhances the surface electronic continuum through resonance effect.

\begin{acknowledgments}
G.B. and H.-H.K acknowledge support from the U.S. DOE, BES grant DE-SC0005463 for spectroscopic studies.
S.O., M.S., N.K. and M.B. acknowledge support by Gordon and Betty Moore Foundation's EPiQS initiative (GBMF4418) and NSF(EFMA-1542798) for film growth.
S.-W.C. and X.W. acknowledge support from NSF Award DMREF-1233349 for single crystal growth.
G.B. also acknowledges partial support from QuantEmX grant from ICAM and the Gordon and Betty Moore Foundation through Grant GBMF5305 and from the
European Regional Development Fund project TK134.
\end{acknowledgments}
\appendix*
\section{Removal of polarization leakage}
\begin{figure}[h]
	\includegraphics[width=6.5cm]{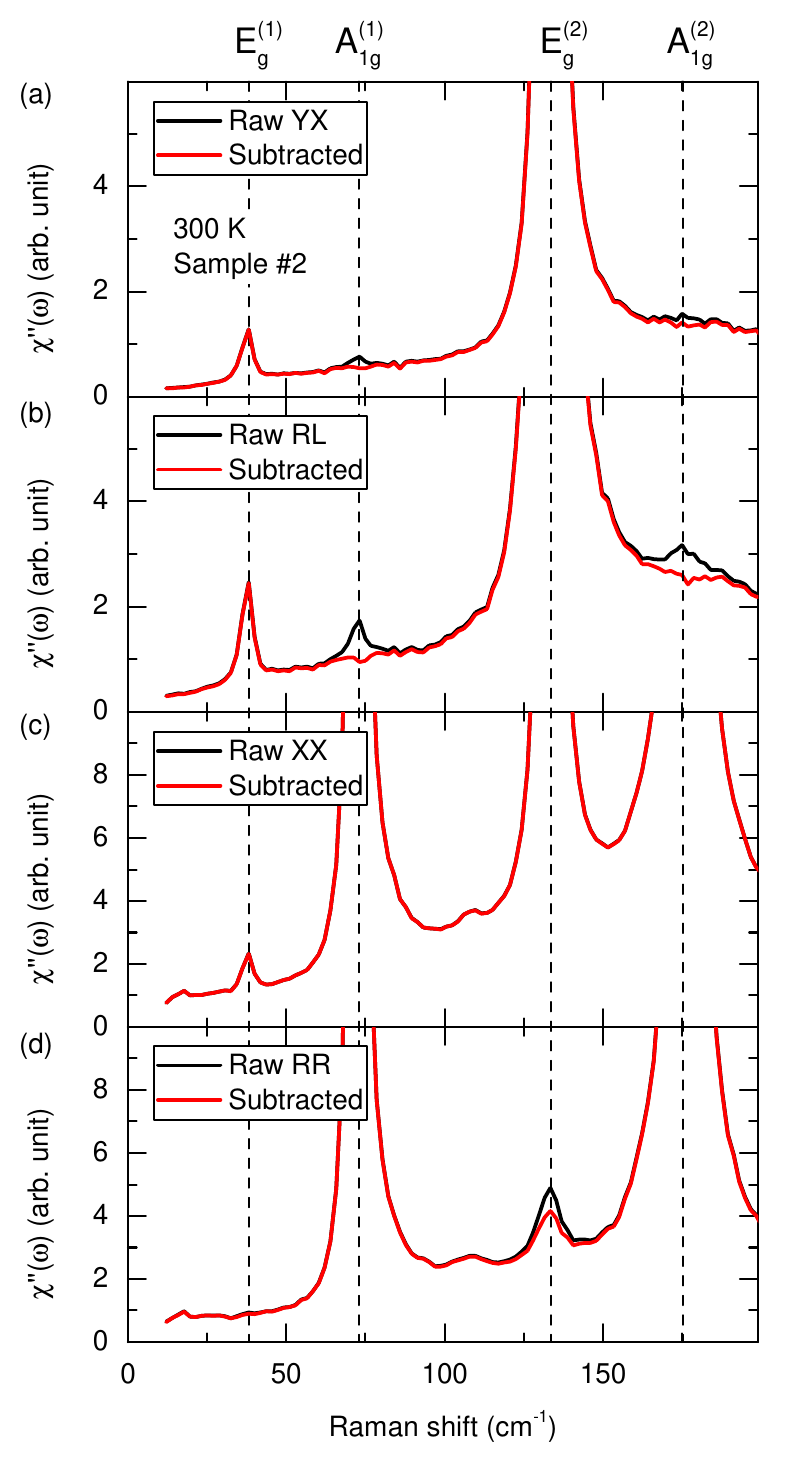}
	\caption{\label{fig:S1}(Color online)
		Comparison of raw data and polarization leakage removed spectra, taken in (a) YX, (b) RL, (c) XX, and (d) RR polarization geometry from the \textit{ab} surface of Sample\,\#2 at 300\,K, with 532\,nm excitation.
	}
\end{figure}
In this section, we explain the details of data analysis concerning removal of polarization leakage from optical elements.
The degree of leakage are determined from the $A_{1g}^{(1)}$ and $A_{1g}^{(2)}$ bulk phonons of single crystal samples at room temperature.
The removal of polarization leakage is done by subtracting intensity from the orthogonal polarization geometry, i.e., $\chi_{\text{YX}}^{\prime\prime}(\omega)=\overline{\chi_{\text{YX}}^{\prime\prime}}(\omega)-\alpha\cdot\overline{\chi_{\text{XX}}^{\prime\prime}}(\omega)$, where $\overline{\chi_{\text{YX}}^{\prime\prime}}(\omega)$ and $\overline{\chi_{\text{XX}}^{\prime\prime}}(\omega)$ are raw data taken in YX and XX polarization geometries, respectively, and $\alpha$ is the leakage ratio due to the limitations of polarization optics. 
It is reasonable to suggest that the same ratio also applies to XX polarization geometry: 
$\chi_{\text{XX}}^{\prime\prime}(\omega)=\overline{\chi_{\text{XX}}^{\prime\prime}}(\omega)-\alpha\cdot\overline{\chi_{\text{YX}}^{\prime\prime}}(\omega)$.
Similarly, we have $\chi_{\text{RL}}^{\prime\prime}(\omega)=\overline{\chi_{\text{RL}}^{\prime\prime}}(\omega)-\beta\cdot\overline{\chi_{\text{RR}}^{\prime\prime}}(\omega)$ and $\chi_{\text{RR}}^{\prime\prime}(\omega)=\overline{\chi_{\text{RR}}^{\prime\prime}}(\omega)-\beta\cdot\overline{\chi_{\text{RL}}^{\prime\prime}}(\omega)$ for the circularly polarized geometries, where $\beta$ is the leakage ratio due to the limitations of the broadband quarter wave plate and alignment of the Berek compensator. 
The ratios $\alpha$ and $\beta$ are in general a weak function of $\omega$, but in a narrow energy window as in this study, they can be safely assumed as constants.
In order to avoid confusion from contributions of surface phonons, we chose YX and RL geometries as our reference for determination of $\alpha$ and $\beta$. 
In these two geometries, only $E_g^{(1)}$ and $E_g^{(2)}$ bulk phonons are expected to be present, the $E$ symmetry surface modes are extremely weak and close to the bulk phonons [Fig.~\ref{fig:hiRes}], and therefore do not raise concern for determination of $\alpha$ and $\beta$.

In Fig.~\ref{fig:S1}, we show spectra of unprocessed raw data and polarization leakage removed results taken at 300\,K from the \textit{ab} surface of a Bi$_2$Se$_3$ thick film in black and red lines, respectively. 
The leakage intensity of $A_{1g}^{(1)}$ and $A_{1g}^{(2)}$ bulk phonons in raw data taken with YX and RL geometries can be fully removed with leakage ratios $\alpha=0.004$ and $\beta=0.015$, respectively. 
These values are within the specification of used broadband polarization optics.

The value of $\alpha$ depends only on the wavelength of light, and therefore the same value $\alpha=0.004$ is used for all samples and temperatures measured with 532\,nm excitation. 
The value of $\beta$ depends critically on the alignment of the Berek compensator, which may vary between experiments, and has to be determined using the method described above in each experiment. 
In this study, the value of $\beta$ is always within the range $0.015\pm 0.005$.
%
%\bibliography{../../Kung2017}

\begin{thebibliography}{77}%
	\makeatletter
	\providecommand \@ifxundefined [1]{%
		\@ifx{#1\undefined}
	}%
	\providecommand \@ifnum [1]{%
		\ifnum #1\expandafter \@firstoftwo
		\else \expandafter \@secondoftwo
		\fi
	}%
	\providecommand \@ifx [1]{%
		\ifx #1\expandafter \@firstoftwo
		\else \expandafter \@secondoftwo
		\fi
	}%
	\providecommand \natexlab [1]{#1}%
	\providecommand \enquote  [1]{``#1''}%
	\providecommand \bibnamefont  [1]{#1}%
	\providecommand \bibfnamefont [1]{#1}%
	\providecommand \citenamefont [1]{#1}%
	\providecommand \href@noop [0]{\@secondoftwo}%
	\providecommand \href [0]{\begingroup \@sanitize@url \@href}%
	\providecommand \@href[1]{\@@startlink{#1}\@@href}%
	\providecommand \@@href[1]{\endgroup#1\@@endlink}%
	\providecommand \@sanitize@url [0]{\catcode `\\12\catcode `\$12\catcode
		`\&12\catcode `\#12\catcode `\^12\catcode `\_12\catcode `\%12\relax}%
	\providecommand \@@startlink[1]{}%
	\providecommand \@@endlink[0]{}%
	\providecommand \url  [0]{\begingroup\@sanitize@url \@url }%
	\providecommand \@url [1]{\endgroup\@href {#1}{\urlprefix }}%
	\providecommand \urlprefix  [0]{URL }%
	\providecommand \Eprint [0]{\href }%
	\providecommand \doibase [0]{http://dx.doi.org/}%
	\providecommand \selectlanguage [0]{\@gobble}%
	\providecommand \bibinfo  [0]{\@secondoftwo}%
	\providecommand \bibfield  [0]{\@secondoftwo}%
	\providecommand \translation [1]{[#1]}%
	\providecommand \BibitemOpen [0]{}%
	\providecommand \bibitemStop [0]{}%
	\providecommand \bibitemNoStop [0]{.\EOS\space}%
	\providecommand \EOS [0]{\spacefactor3000\relax}%
	\providecommand \BibitemShut  [1]{\csname bibitem#1\endcsname}%
	\let\auto@bib@innerbib\@empty
	%</preamble>
	\bibitem [{\citenamefont {Fu}\ \emph {et~al.}(2007)\citenamefont {Fu},
		\citenamefont {Kane},\ and\ \citenamefont {Mele}}]{Fu2007A}%
	\BibitemOpen
	\bibfield  {author} {\bibinfo {author} {\bibfnamefont {Liang}\ \bibnamefont
			{Fu}}, \bibinfo {author} {\bibfnamefont {C.~L.}\ \bibnamefont {Kane}}, \ and\
		\bibinfo {author} {\bibfnamefont {E.~J.}\ \bibnamefont {Mele}},\ }\bibfield
	{title} {\enquote {\bibinfo {title} {Topological insulators in three
				dimensions},}\ }\href {\doibase 10.1103/PhysRevLett.98.106803} {\bibfield
		{journal} {\bibinfo  {journal} {Phys. Rev. Lett.}\ }\textbf {\bibinfo
			{volume} {98}},\ \bibinfo {pages} {106803} (\bibinfo {year}
		{2007})}\BibitemShut {NoStop}%
	\bibitem [{\citenamefont {Zhang}\ \emph {et~al.}(2009)\citenamefont {Zhang},
		\citenamefont {Liu}, \citenamefont {Qi}, \citenamefont {Dai}, \citenamefont
		{Fang},\ and\ \citenamefont {Zhang}}]{Zhang2009}%
	\BibitemOpen
	\bibfield  {author} {\bibinfo {author} {\bibfnamefont {Haijun}\ \bibnamefont
			{Zhang}}, \bibinfo {author} {\bibfnamefont {Chao-Xing}\ \bibnamefont {Liu}},
		\bibinfo {author} {\bibfnamefont {Xiao-Liang}\ \bibnamefont {Qi}}, \bibinfo
		{author} {\bibfnamefont {Xi}~\bibnamefont {Dai}}, \bibinfo {author}
		{\bibfnamefont {Zhong}\ \bibnamefont {Fang}}, \ and\ \bibinfo {author}
		{\bibfnamefont {Shou-Cheng}\ \bibnamefont {Zhang}},\ }\bibfield  {title}
	{\enquote {\bibinfo {title} {Topological insulators in {Bi$_{2}$Se$_{3}$},
				{Bi$_{2}$Te$_{3}$} and {Sb$_{2}$Te$_{3}$} with a single {D}irac cone on the
				surface},}\ }\href {\doibase 10.1038/nphys1270} {\bibfield  {journal}
		{\bibinfo  {journal} {Nature Phys.}\ }\textbf {\bibinfo {volume} {5}},\
		\bibinfo {pages} {438} (\bibinfo {year} {2009})}\BibitemShut {NoStop}%
	\bibitem [{\citenamefont {Hsieh}\ \emph {et~al.}(2009)\citenamefont {Hsieh},
		\citenamefont {Xia}, \citenamefont {Qian}, \citenamefont {Wray},
		\citenamefont {Dil}, \citenamefont {Meier}, \citenamefont {Osterwalder},
		\citenamefont {Patthey}, \citenamefont {Checkelsky}, \citenamefont {Ong},
		\citenamefont {Fedorov}, \citenamefont {Lin}, \citenamefont {Bansil},
		\citenamefont {Grauer}, \citenamefont {Hor}, \citenamefont {Cava},\ and\
		\citenamefont {Hasan}}]{Hsieh2009}%
	\BibitemOpen
	\bibfield  {author} {\bibinfo {author} {\bibfnamefont {D.}~\bibnamefont
			{Hsieh}}, \bibinfo {author} {\bibfnamefont {Y.}~\bibnamefont {Xia}}, \bibinfo
		{author} {\bibfnamefont {D.}~\bibnamefont {Qian}}, \bibinfo {author}
		{\bibfnamefont {L.}~\bibnamefont {Wray}}, \bibinfo {author} {\bibfnamefont
			{J.~H.}\ \bibnamefont {Dil}}, \bibinfo {author} {\bibfnamefont
			{F.}~\bibnamefont {Meier}}, \bibinfo {author} {\bibfnamefont
			{J.}~\bibnamefont {Osterwalder}}, \bibinfo {author} {\bibfnamefont
			{L.}~\bibnamefont {Patthey}}, \bibinfo {author} {\bibfnamefont {J.~G.}\
			\bibnamefont {Checkelsky}}, \bibinfo {author} {\bibfnamefont {N.~P.}\
			\bibnamefont {Ong}}, \bibinfo {author} {\bibfnamefont {A.~V.}\ \bibnamefont
			{Fedorov}}, \bibinfo {author} {\bibfnamefont {H.}~\bibnamefont {Lin}},
		\bibinfo {author} {\bibfnamefont {A.}~\bibnamefont {Bansil}}, \bibinfo
		{author} {\bibfnamefont {D.}~\bibnamefont {Grauer}}, \bibinfo {author}
		{\bibfnamefont {Y.~S.}\ \bibnamefont {Hor}}, \bibinfo {author} {\bibfnamefont
			{R.~J.}\ \bibnamefont {Cava}}, \ and\ \bibinfo {author} {\bibfnamefont
			{M.~Z.}\ \bibnamefont {Hasan}},\ }\bibfield  {title} {\enquote {\bibinfo
			{title} {A tunable topological insulator in the spin helical {D}irac
				transport regime},}\ }\href {\doibase 10.1038/nature08234} {\bibfield
		{journal} {\bibinfo  {journal} {Nature}\ }\textbf {\bibinfo {volume} {460}},\
		\bibinfo {pages} {1101--1105} (\bibinfo {year} {2009})}\BibitemShut {NoStop}%
	\bibitem [{\citenamefont {Xia}\ \emph {et~al.}(2009)\citenamefont {Xia},
		\citenamefont {Qian}, \citenamefont {Hsieh}, \citenamefont {Wray},
		\citenamefont {Pal}, \citenamefont {Lin}, \citenamefont {Bansil},
		\citenamefont {Grauer}, \citenamefont {Hor}, \citenamefont {Cava},\ and\
		\citenamefont {Hasan}}]{Xia2009}%
	\BibitemOpen
	\bibfield  {author} {\bibinfo {author} {\bibfnamefont {Y.}~\bibnamefont
			{Xia}}, \bibinfo {author} {\bibfnamefont {D.}~\bibnamefont {Qian}}, \bibinfo
		{author} {\bibfnamefont {D.}~\bibnamefont {Hsieh}}, \bibinfo {author}
		{\bibfnamefont {L.}~\bibnamefont {Wray}}, \bibinfo {author} {\bibfnamefont
			{A.}~\bibnamefont {Pal}}, \bibinfo {author} {\bibfnamefont {H.}~\bibnamefont
			{Lin}}, \bibinfo {author} {\bibfnamefont {A.}~\bibnamefont {Bansil}},
		\bibinfo {author} {\bibfnamefont {D.}~\bibnamefont {Grauer}}, \bibinfo
		{author} {\bibfnamefont {Y.~S.}\ \bibnamefont {Hor}}, \bibinfo {author}
		{\bibfnamefont {R.~J.}\ \bibnamefont {Cava}}, \ and\ \bibinfo {author}
		{\bibfnamefont {M.~Z.}\ \bibnamefont {Hasan}},\ }\bibfield  {title} {\enquote
		{\bibinfo {title} {Observation of a large-gap topological-insulator class
				with a single {D}irac cone on the surface},}\ }\href {\doibase
		10.1038/nphys1274} {\bibfield  {journal} {\bibinfo  {journal} {Nature Phys.}\
		}\textbf {\bibinfo {volume} {5}},\ \bibinfo {pages} {398--402} (\bibinfo
		{year} {2009})}\BibitemShut {NoStop}%
	\bibitem [{\citenamefont {Checkelsky}\ \emph {et~al.}(2009)\citenamefont
		{Checkelsky}, \citenamefont {Hor}, \citenamefont {Liu}, \citenamefont {Qu},
		\citenamefont {Cava},\ and\ \citenamefont {Ong}}]{Checkelsky2009}%
	\BibitemOpen
	\bibfield  {author} {\bibinfo {author} {\bibfnamefont {J.~G.}\ \bibnamefont
			{Checkelsky}}, \bibinfo {author} {\bibfnamefont {Y.~S.}\ \bibnamefont {Hor}},
		\bibinfo {author} {\bibfnamefont {M.-H.}\ \bibnamefont {Liu}}, \bibinfo
		{author} {\bibfnamefont {D.-X.}\ \bibnamefont {Qu}}, \bibinfo {author}
		{\bibfnamefont {R.~J.}\ \bibnamefont {Cava}}, \ and\ \bibinfo {author}
		{\bibfnamefont {N.~P.}\ \bibnamefont {Ong}},\ }\bibfield  {title} {\enquote
		{\bibinfo {title} {Quantum interference in macroscopic crystals of
				nonmetallic {Bi}$_2${Se}$_3$},}\ }\href {\doibase
		10.1103/PhysRevLett.103.246601} {\bibfield  {journal} {\bibinfo  {journal}
			{Phys. Rev. Lett.}\ }\textbf {\bibinfo {volume} {103}},\ \bibinfo {pages}
		{246601} (\bibinfo {year} {2009})}\BibitemShut {NoStop}%
	\bibitem [{\citenamefont {Bianchi}\ \emph {et~al.}(2010)\citenamefont
		{Bianchi}, \citenamefont {Guan}, \citenamefont {Bao}, \citenamefont {Mi},
		\citenamefont {Iversen}, \citenamefont {King},\ and\ \citenamefont
		{Hofmann}}]{Bianchi2010}%
	\BibitemOpen
	\bibfield  {author} {\bibinfo {author} {\bibfnamefont {Marco}\ \bibnamefont
			{Bianchi}}, \bibinfo {author} {\bibfnamefont {Dandan}\ \bibnamefont {Guan}},
		\bibinfo {author} {\bibfnamefont {Shining}\ \bibnamefont {Bao}}, \bibinfo
		{author} {\bibfnamefont {Jianli}\ \bibnamefont {Mi}}, \bibinfo {author}
		{\bibfnamefont {Bo~Brummerstedt}\ \bibnamefont {Iversen}}, \bibinfo {author}
		{\bibfnamefont {Philip D.~C.}\ \bibnamefont {King}}, \ and\ \bibinfo {author}
		{\bibfnamefont {Philip}\ \bibnamefont {Hofmann}},\ }\bibfield  {title}
	{\enquote {\bibinfo {title} {Coexistence of the topological state and a
				two-dimensional electron gas on the surface of {Bi$_{2}$Se$_{3}$}},}\ }\href
	{http://dx.doi.org/10.1038/ncomms1131} {\bibfield  {journal} {\bibinfo
			{journal} {Nat. Commun.}\ }\textbf {\bibinfo {volume} {1}},\ \bibinfo {pages}
		{128} (\bibinfo {year} {2010})}\BibitemShut {NoStop}%
	\bibitem [{\citenamefont {Beidenkopf}\ \emph {et~al.}(2011)\citenamefont
		{Beidenkopf}, \citenamefont {Roushan}, \citenamefont {Seo}, \citenamefont
		{Gorman}, \citenamefont {Drozdov}, \citenamefont {Hor}, \citenamefont
		{Cava},\ and\ \citenamefont {Yazdani}}]{Beidenkopf2011}%
	\BibitemOpen
	\bibfield  {author} {\bibinfo {author} {\bibfnamefont {Haim}\ \bibnamefont
			{Beidenkopf}}, \bibinfo {author} {\bibfnamefont {Pedram}\ \bibnamefont
			{Roushan}}, \bibinfo {author} {\bibfnamefont {Jungpil}\ \bibnamefont {Seo}},
		\bibinfo {author} {\bibfnamefont {Lindsay}\ \bibnamefont {Gorman}}, \bibinfo
		{author} {\bibfnamefont {Ilya}\ \bibnamefont {Drozdov}}, \bibinfo {author}
		{\bibfnamefont {Yew~San}\ \bibnamefont {Hor}}, \bibinfo {author}
		{\bibfnamefont {R.~J.}\ \bibnamefont {Cava}}, \ and\ \bibinfo {author}
		{\bibfnamefont {Ali}\ \bibnamefont {Yazdani}},\ }\bibfield  {title} {\enquote
		{\bibinfo {title} {Spatial fluctuations of helical {D}irac fermions on the
				surface of topological insulators},}\ }\href {\doibase 10.1038/nphys2108}
	{\bibfield  {journal} {\bibinfo  {journal} {Nature Phys.}\ }\textbf {\bibinfo
			{volume} {7}},\ \bibinfo {pages} {939--943} (\bibinfo {year}
		{2011})}\BibitemShut {NoStop}%
	\bibitem [{\citenamefont {Hasan}\ and\ \citenamefont
		{Moore}(2011)}]{Hasan2011}%
	\BibitemOpen
	\bibfield  {author} {\bibinfo {author} {\bibfnamefont {M.~Zahid}\
			\bibnamefont {Hasan}}\ and\ \bibinfo {author} {\bibfnamefont {Joel~E.}\
			\bibnamefont {Moore}},\ }\bibfield  {title} {\enquote {\bibinfo {title}
			{Three-dimensional topological insulators},}\ }\href {\doibase
		10.1146/annurev-conmatphys-062910-140432} {\bibfield  {journal} {\bibinfo
			{journal} {Annual Review of Condensed Matter Physics}\ }\textbf {\bibinfo
			{volume} {2}},\ \bibinfo {pages} {55--78} (\bibinfo {year}
		{2011})}\BibitemShut {NoStop}%
	\bibitem [{\citenamefont {Fu}\ and\ \citenamefont {Kane}(2008)}]{Fu2008}%
	\BibitemOpen
	\bibfield  {author} {\bibinfo {author} {\bibfnamefont {Liang}\ \bibnamefont
			{Fu}}\ and\ \bibinfo {author} {\bibfnamefont {C.~L.}\ \bibnamefont {Kane}},\
	}\bibfield  {title} {\enquote {\bibinfo {title} {Superconducting proximity
			effect and majorana fermions at the surface of a topological insulator},}\
}\href {\doibase 10.1103/PhysRevLett.100.096407} {\bibfield  {journal}
{\bibinfo  {journal} {Phys. Rev. Lett.}\ }\textbf {\bibinfo {volume} {100}},\
\bibinfo {pages} {096407} (\bibinfo {year} {2008})}\BibitemShut {NoStop}%
\bibitem [{\citenamefont {Qi}\ \emph {et~al.}(2008)\citenamefont {Qi},
	\citenamefont {Hughes},\ and\ \citenamefont {Zhang}}]{Qi2008}%
\BibitemOpen
\bibfield  {author} {\bibinfo {author} {\bibfnamefont {Xiao-Liang}\
		\bibnamefont {Qi}}, \bibinfo {author} {\bibfnamefont {Taylor~L.}\
		\bibnamefont {Hughes}}, \ and\ \bibinfo {author} {\bibfnamefont {Shou-Cheng}\
		\bibnamefont {Zhang}},\ }\bibfield  {title} {\enquote {\bibinfo {title}
		{Topological field theory of time-reversal invariant insulators},}\ }\href
{\doibase 10.1103/PhysRevB.78.195424} {\bibfield  {journal} {\bibinfo
		{journal} {Phys. Rev. B}\ }\textbf {\bibinfo {volume} {78}},\ \bibinfo
	{pages} {195424} (\bibinfo {year} {2008})}\BibitemShut {NoStop}%
\bibitem [{\citenamefont {Qi}\ \emph {et~al.}(2009)\citenamefont {Qi},
	\citenamefont {Li}, \citenamefont {Zang},\ and\ \citenamefont
	{Zhang}}]{Qi2009}%
\BibitemOpen
\bibfield  {author} {\bibinfo {author} {\bibfnamefont {Xiao-Liang}\
		\bibnamefont {Qi}}, \bibinfo {author} {\bibfnamefont {Rundong}\ \bibnamefont
		{Li}}, \bibinfo {author} {\bibfnamefont {Jiadong}\ \bibnamefont {Zang}}, \
	and\ \bibinfo {author} {\bibfnamefont {Shou-Cheng}\ \bibnamefont {Zhang}},\
}\bibfield  {title} {\enquote {\bibinfo {title} {Inducing a magnetic monopole
		with topological surface states},}\ }\href {\doibase 10.1126/science.1167747}
{\bibfield  {journal} {\bibinfo  {journal} {Science}\ }\textbf {\bibinfo
		{volume} {323}},\ \bibinfo {pages} {1184--1187} (\bibinfo {year}
	{2009})}\BibitemShut {NoStop}%
\bibitem [{\citenamefont {Hasan}\ and\ \citenamefont {Kane}(2010)}]{Hasan2010}%
\BibitemOpen
\bibfield  {author} {\bibinfo {author} {\bibfnamefont {M.~Z.}\ \bibnamefont
		{Hasan}}\ and\ \bibinfo {author} {\bibfnamefont {C.~L.}\ \bibnamefont
		{Kane}},\ }\bibfield  {title} {\enquote {\bibinfo {title}
		{\textit{Colloquium} : Topological insulators},}\ }\href {\doibase
	10.1103/RevModPhys.82.3045} {\bibfield  {journal} {\bibinfo  {journal} {Rev.
			Mod. Phys.}\ }\textbf {\bibinfo {volume} {82}},\ \bibinfo {pages}
	{3045--3067} (\bibinfo {year} {2010})}\BibitemShut {NoStop}%
\bibitem [{\citenamefont {Yu}\ \emph {et~al.}(2010)\citenamefont {Yu},
	\citenamefont {Zhang}, \citenamefont {Zhang}, \citenamefont {Zhang},
	\citenamefont {Dai},\ and\ \citenamefont {Fang}}]{Yu2010}%
\BibitemOpen
\bibfield  {author} {\bibinfo {author} {\bibfnamefont {Rui}\ \bibnamefont
		{Yu}}, \bibinfo {author} {\bibfnamefont {Wei}\ \bibnamefont {Zhang}},
	\bibinfo {author} {\bibfnamefont {Hai-Jun}\ \bibnamefont {Zhang}}, \bibinfo
	{author} {\bibfnamefont {Shou-Cheng}\ \bibnamefont {Zhang}}, \bibinfo
	{author} {\bibfnamefont {Xi}~\bibnamefont {Dai}}, \ and\ \bibinfo {author}
	{\bibfnamefont {Zhong}\ \bibnamefont {Fang}},\ }\bibfield  {title} {\enquote
	{\bibinfo {title} {Quantized anomalous hall effect in magnetic topological
			insulators},}\ }\href {\doibase 10.1126/science.1187485} {\bibfield
	{journal} {\bibinfo  {journal} {Science}\ }\textbf {\bibinfo {volume}
		{329}},\ \bibinfo {pages} {61--64} (\bibinfo {year} {2010})}\BibitemShut
{NoStop}%
\bibitem [{\citenamefont {Raghu}\ \emph {et~al.}(2010)\citenamefont {Raghu},
	\citenamefont {Chung}, \citenamefont {Qi},\ and\ \citenamefont
	{Zhang}}]{Raghu2010}%
\BibitemOpen
\bibfield  {author} {\bibinfo {author} {\bibfnamefont {S.}~\bibnamefont
		{Raghu}}, \bibinfo {author} {\bibfnamefont {Suk~Bum}\ \bibnamefont {Chung}},
	\bibinfo {author} {\bibfnamefont {Xiao-Liang}\ \bibnamefont {Qi}}, \ and\
	\bibinfo {author} {\bibfnamefont {Shou-Cheng}\ \bibnamefont {Zhang}},\
}\bibfield  {title} {\enquote {\bibinfo {title} {Collective modes of a
		helical liquid},}\ }\href {\doibase 10.1103/PhysRevLett.104.116401}
{\bibfield  {journal} {\bibinfo  {journal} {Phys. Rev. Lett.}\ }\textbf
	{\bibinfo {volume} {104}},\ \bibinfo {pages} {116401} (\bibinfo {year}
	{2010})}\BibitemShut {NoStop}%
\bibitem [{\citenamefont {Qi}\ and\ \citenamefont {Zhang}(2011)}]{Qi2011}%
\BibitemOpen
\bibfield  {author} {\bibinfo {author} {\bibfnamefont {Xiao-Liang}\
		\bibnamefont {Qi}}\ and\ \bibinfo {author} {\bibfnamefont {Shou-Cheng}\
		\bibnamefont {Zhang}},\ }\bibfield  {title} {\enquote {\bibinfo {title}
		{Topological insulators and superconductors},}\ }\href {\doibase
	10.1103/RevModPhys.83.1057} {\bibfield  {journal} {\bibinfo  {journal} {Rev.
			Mod. Phys.}\ }\textbf {\bibinfo {volume} {83}},\ \bibinfo {pages}
	{1057--1110} (\bibinfo {year} {2011})}\BibitemShut {NoStop}%
\bibitem [{\citenamefont {Wang}\ \emph {et~al.}(2013)\citenamefont {Wang},
	\citenamefont {Steinberg}, \citenamefont {Jarillo-Herrero},\ and\
	\citenamefont {Gedik}}]{Wang2013B}%
\BibitemOpen
\bibfield  {author} {\bibinfo {author} {\bibfnamefont {Y.~H.}\ \bibnamefont
		{Wang}}, \bibinfo {author} {\bibfnamefont {H.}~\bibnamefont {Steinberg}},
	\bibinfo {author} {\bibfnamefont {P.}~\bibnamefont {Jarillo-Herrero}}, \ and\
	\bibinfo {author} {\bibfnamefont {N.}~\bibnamefont {Gedik}},\ }\bibfield
{title} {\enquote {\bibinfo {title} {Observation of {F}loquet-{B}loch states
			on the surface of a topological insulator},}\ }\href {\doibase
	10.1126/science.1239834} {\bibfield  {journal} {\bibinfo  {journal}
		{Science}\ }\textbf {\bibinfo {volume} {342}},\ \bibinfo {pages} {453--457}
	(\bibinfo {year} {2013})}\BibitemShut {NoStop}%
\bibitem [{\citenamefont {Grover}\ \emph {et~al.}(2014)\citenamefont {Grover},
	\citenamefont {Sheng},\ and\ \citenamefont {Vishwanath}}]{Grover2014}%
\BibitemOpen
\bibfield  {author} {\bibinfo {author} {\bibfnamefont {Tarun}\ \bibnamefont
		{Grover}}, \bibinfo {author} {\bibfnamefont {D.~N.}\ \bibnamefont {Sheng}}, \
	and\ \bibinfo {author} {\bibfnamefont {Ashvin}\ \bibnamefont {Vishwanath}},\
}\bibfield  {title} {\enquote {\bibinfo {title} {Emergent space-time
		supersymmetry at the boundary of a topological phase},}\ }\href {\doibase
10.1126/science.1248253} {\bibfield  {journal} {\bibinfo  {journal}
	{Science}\ }\textbf {\bibinfo {volume} {344}},\ \bibinfo {pages} {280--283}
(\bibinfo {year} {2014})}\BibitemShut {NoStop}%
\bibitem [{\citenamefont {Butch}\ \emph {et~al.}(2010)\citenamefont {Butch},
	\citenamefont {Kirshenbaum}, \citenamefont {Syers}, \citenamefont {Sushkov},
	\citenamefont {Jenkins}, \citenamefont {Drew},\ and\ \citenamefont
	{Paglione}}]{Butch2010B}%
\BibitemOpen
\bibfield  {author} {\bibinfo {author} {\bibfnamefont {N.~P.}\ \bibnamefont
		{Butch}}, \bibinfo {author} {\bibfnamefont {K.}~\bibnamefont {Kirshenbaum}},
	\bibinfo {author} {\bibfnamefont {P.}~\bibnamefont {Syers}}, \bibinfo
	{author} {\bibfnamefont {A.~B.}\ \bibnamefont {Sushkov}}, \bibinfo {author}
	{\bibfnamefont {G.~S.}\ \bibnamefont {Jenkins}}, \bibinfo {author}
	{\bibfnamefont {H.~D.}\ \bibnamefont {Drew}}, \ and\ \bibinfo {author}
	{\bibfnamefont {J.}~\bibnamefont {Paglione}},\ }\bibfield  {title} {\enquote
	{\bibinfo {title} {Strong surface scattering in ultrahigh-mobility
			{Bi}$_2${Se}$_3$ topological insulator crystals},}\ }\href {\doibase
	10.1103/PhysRevB.81.241301} {\bibfield  {journal} {\bibinfo  {journal} {Phys.
			Rev. B}\ }\textbf {\bibinfo {volume} {81}},\ \bibinfo {pages} {241301}
	(\bibinfo {year} {2010})}\BibitemShut {NoStop}%
\bibitem [{\citenamefont {Pan}\ \emph {et~al.}(2012)\citenamefont {Pan},
	\citenamefont {Fedorov}, \citenamefont {Gardner}, \citenamefont {Lee},
	\citenamefont {Chu},\ and\ \citenamefont {Valla}}]{Pan2012}%
\BibitemOpen
\bibfield  {author} {\bibinfo {author} {\bibfnamefont {Z.-H.}\ \bibnamefont
		{Pan}}, \bibinfo {author} {\bibfnamefont {A.~V.}\ \bibnamefont {Fedorov}},
	\bibinfo {author} {\bibfnamefont {D.}~\bibnamefont {Gardner}}, \bibinfo
	{author} {\bibfnamefont {Y.~S.}\ \bibnamefont {Lee}}, \bibinfo {author}
	{\bibfnamefont {S.}~\bibnamefont {Chu}}, \ and\ \bibinfo {author}
	{\bibfnamefont {T.}~\bibnamefont {Valla}},\ }\bibfield  {title} {\enquote
	{\bibinfo {title} {Measurement of an exceptionally weak electron-phonon
			coupling on the surface of the topological insulator {Bi}$_2${Se}$_3$ using
			angle-resolved photoemission spectroscopy},}\ }\href {\doibase
	10.1103/PhysRevLett.108.187001} {\bibfield  {journal} {\bibinfo  {journal}
		{Phys. Rev. Lett.}\ }\textbf {\bibinfo {volume} {108}},\ \bibinfo {pages}
	{187001} (\bibinfo {year} {2012})}\BibitemShut {NoStop}%
\bibitem [{\citenamefont {Valla}\ \emph {et~al.}(2012)\citenamefont {Valla},
	\citenamefont {Pan}, \citenamefont {Gardner}, \citenamefont {Lee},\ and\
	\citenamefont {Chu}}]{Valla2012}%
\BibitemOpen
\bibfield  {author} {\bibinfo {author} {\bibfnamefont {T.}~\bibnamefont
		{Valla}}, \bibinfo {author} {\bibfnamefont {Z.-H.}\ \bibnamefont {Pan}},
	\bibinfo {author} {\bibfnamefont {D.}~\bibnamefont {Gardner}}, \bibinfo
	{author} {\bibfnamefont {Y.~S.}\ \bibnamefont {Lee}}, \ and\ \bibinfo
	{author} {\bibfnamefont {S.}~\bibnamefont {Chu}},\ }\bibfield  {title}
{\enquote {\bibinfo {title} {Photoemission spectroscopy of magnetic and
			nonmagnetic impurities on the surface of the {Bi}$_2${Se}$_3$ topological
			insulator},}\ }\href {\doibase 10.1103/PhysRevLett.108.117601} {\bibfield
	{journal} {\bibinfo  {journal} {Phys. Rev. Lett.}\ }\textbf {\bibinfo
		{volume} {108}},\ \bibinfo {pages} {117601} (\bibinfo {year}
	{2012})}\BibitemShut {NoStop}%
\bibitem [{\citenamefont {Parente}\ \emph {et~al.}(2013)\citenamefont
	{Parente}, \citenamefont {Tagliacozzo}, \citenamefont {von Oppen},\ and\
	\citenamefont {Guinea}}]{Parente2013}%
\BibitemOpen
\bibfield  {author} {\bibinfo {author} {\bibfnamefont {V.}~\bibnamefont
		{Parente}}, \bibinfo {author} {\bibfnamefont {A.}~\bibnamefont
		{Tagliacozzo}}, \bibinfo {author} {\bibfnamefont {F.}~\bibnamefont {von
			Oppen}}, \ and\ \bibinfo {author} {\bibfnamefont {F.}~\bibnamefont
		{Guinea}},\ }\bibfield  {title} {\enquote {\bibinfo {title} {Electron-phonon
			interaction on the surface of a three-dimensional topological insulator},}\
}\href {\doibase 10.1103/PhysRevB.88.075432} {\bibfield  {journal} {\bibinfo
	{journal} {Phys. Rev. B}\ }\textbf {\bibinfo {volume} {88}},\ \bibinfo
{pages} {075432} (\bibinfo {year} {2013})}\BibitemShut {NoStop}%
\bibitem [{\citenamefont {Costache}\ \emph {et~al.}(2014)\citenamefont
	{Costache}, \citenamefont {Neumann}, \citenamefont {Sierra}, \citenamefont
	{Marinova}, \citenamefont {Gospodinov}, \citenamefont {Roche},\ and\
	\citenamefont {Valenzuela}}]{Costache2014}%
\BibitemOpen
\bibfield  {author} {\bibinfo {author} {\bibfnamefont {M.~V.}\ \bibnamefont
		{Costache}}, \bibinfo {author} {\bibfnamefont {I.}~\bibnamefont {Neumann}},
	\bibinfo {author} {\bibfnamefont {J.~F.}\ \bibnamefont {Sierra}}, \bibinfo
	{author} {\bibfnamefont {V.}~\bibnamefont {Marinova}}, \bibinfo {author}
	{\bibfnamefont {M.~M.}\ \bibnamefont {Gospodinov}}, \bibinfo {author}
	{\bibfnamefont {S.}~\bibnamefont {Roche}}, \ and\ \bibinfo {author}
	{\bibfnamefont {S.~O.}\ \bibnamefont {Valenzuela}},\ }\bibfield  {title}
{\enquote {\bibinfo {title} {Fingerprints of inelastic transport at the
			surface of the topological insulator {Bi}$_2${Se}$_3$: Role of
			electron-phonon coupling},}\ }\href {\doibase 10.1103/PhysRevLett.112.086601}
{\bibfield  {journal} {\bibinfo  {journal} {Phys. Rev. Lett.}\ }\textbf
	{\bibinfo {volume} {112}},\ \bibinfo {pages} {086601} (\bibinfo {year}
	{2014})}\BibitemShut {NoStop}%
\bibitem [{\citenamefont {Lifshitz}\ and\ \citenamefont
	{Rosenzweig}(1948)}]{Lifshitz1948}%
\BibitemOpen
\bibfield  {author} {\bibinfo {author} {\bibfnamefont {I.~M.}\ \bibnamefont
		{Lifshitz}}\ and\ \bibinfo {author} {\bibfnamefont {L.~N.}\ \bibnamefont
		{Rosenzweig}},\ }\bibfield  {title} {\enquote {\bibinfo {title} {Dynamics of
			lattice filling half-space ({R}ussian)},}\ }\href@noop {} {\bibfield
	{journal} {\bibinfo  {journal} {Zh. Eksp. Teor. Fiz.}\ }\textbf {\bibinfo
		{volume} {18}},\ \bibinfo {pages} {1012} (\bibinfo {year}
	{1948})}\BibitemShut {NoStop}%
\bibitem [{\citenamefont {Lifshitz}(1956)}]{Lifshitz1956}%
\BibitemOpen
\bibfield  {author} {\bibinfo {author} {\bibfnamefont {I.~M.}\ \bibnamefont
		{Lifshitz}},\ }\bibfield  {title} {\enquote {\bibinfo {title} {Some problems
			of the dynamic theory of non-ideal crystal lattices},}\ }\href {\doibase
	10.1007/BF02746071} {\bibfield  {journal} {\bibinfo  {journal} {Il Nuovo
			Cimento}\ }\textbf {\bibinfo {volume} {3}},\ \bibinfo {pages} {716--734}
	(\bibinfo {year} {1956})}\BibitemShut {NoStop}%
\bibitem [{\citenamefont {Wallis}(1957)}]{Wallis1957}%
\BibitemOpen
\bibfield  {author} {\bibinfo {author} {\bibfnamefont {Richard~F.}\
		\bibnamefont {Wallis}},\ }\bibfield  {title} {\enquote {\bibinfo {title}
		{Effect of free ends on the vibration frequencies of one-dimensional
			lattices},}\ }\href {\doibase 10.1103/PhysRev.105.540} {\bibfield  {journal}
	{\bibinfo  {journal} {Phys. Rev.}\ }\textbf {\bibinfo {volume} {105}},\
	\bibinfo {pages} {540--545} (\bibinfo {year} {1957})}\BibitemShut {NoStop}%
\bibitem [{\citenamefont {Wallis}(1959)}]{Wallis1959}%
\BibitemOpen
\bibfield  {author} {\bibinfo {author} {\bibfnamefont {Richard~F.}\
		\bibnamefont {Wallis}},\ }\bibfield  {title} {\enquote {\bibinfo {title}
		{Theory of surface modes of vibration in two- and three-dimensional crystal
			lattices},}\ }\href {\doibase 10.1103/PhysRev.116.302} {\bibfield  {journal}
	{\bibinfo  {journal} {Phys. Rev.}\ }\textbf {\bibinfo {volume} {116}},\
	\bibinfo {pages} {302--308} (\bibinfo {year} {1959})}\BibitemShut {NoStop}%
\bibitem [{\citenamefont {Benedek}\ and\ \citenamefont
	{Miglio}(1991)}]{Benedek1991}%
\BibitemOpen
\bibfield  {author} {\bibinfo {author} {\bibfnamefont {G.}~\bibnamefont
		{Benedek}}\ and\ \bibinfo {author} {\bibfnamefont {L.}~\bibnamefont
		{Miglio}},\ }\bibfield  {title} {\enquote {\bibinfo {title} {The green's
			function method in the surface lattice dynamics of ionic crystals},}\ }in\
\href {\doibase 10.1007/978-3-642-75785-3_3} {\emph {\bibinfo {booktitle}
		{Surface Phonons}}},\ \bibinfo {editor} {edited by\ \bibinfo {editor}
	{\bibfnamefont {Winfried}\ \bibnamefont {Kress}}\ and\ \bibinfo {editor}
	{\bibfnamefont {Frederik~W.}\ \bibnamefont {de~Wette}}}\ (\bibinfo
{publisher} {Springer Berlin Heidelberg},\ \bibinfo {address} {Berlin,
	Heidelberg},\ \bibinfo {year} {1991})\ pp.\ \bibinfo {pages}
{37--66}\BibitemShut {NoStop}%
\bibitem [{\citenamefont {Wallis}(1994)}]{Wallis1994}%
\BibitemOpen
\bibfield  {author} {\bibinfo {author} {\bibfnamefont {R.F}\ \bibnamefont
		{Wallis}},\ }\bibfield  {title} {\enquote {\bibinfo {title} {Surface phonons:
			theoretical developments},}\ }\href
{http://dx.doi.org/10.1016/0039-6028(94)90684-X} {\bibfield  {journal}
	{\bibinfo  {journal} {Surface Science}\ }\textbf {\bibinfo {volume} {299}},\
	\bibinfo {pages} {612 -- 627} (\bibinfo {year} {1994})}\BibitemShut {NoStop}%
\bibitem [{\citenamefont {Lagos}\ \emph {et~al.}(2017)\citenamefont {Lagos},
	\citenamefont {Tr{\"u}gler}, \citenamefont {Hohenester},\ and\ \citenamefont
	{Batson}}]{Lagos2017}%
\BibitemOpen
\bibfield  {author} {\bibinfo {author} {\bibfnamefont {Maureen~J.}\
		\bibnamefont {Lagos}}, \bibinfo {author} {\bibfnamefont {Andreas}\
		\bibnamefont {Tr{\"u}gler}}, \bibinfo {author} {\bibfnamefont {Ulrich}\
		\bibnamefont {Hohenester}}, \ and\ \bibinfo {author} {\bibfnamefont
		{Philip~E.}\ \bibnamefont {Batson}},\ }\bibfield  {title} {\enquote {\bibinfo
		{title} {Mapping vibrational surface and bulk modes in a single nanocube},}\
}\href {http://dx.doi.org/10.1038/nature21699} {\bibfield  {journal}
{\bibinfo  {journal} {Nature}\ }\textbf {\bibinfo {volume} {543}},\ \bibinfo
{pages} {529--532} (\bibinfo {year} {2017})},\ \bibinfo {note}
{letter}\BibitemShut {NoStop}%
\bibitem [{\citenamefont {Zhao}\ \emph {et~al.}(2011)\citenamefont {Zhao},
	\citenamefont {Beekman}, \citenamefont {Sandilands}, \citenamefont
	{Bashucky}, \citenamefont {Kwok}, \citenamefont {Lee}, \citenamefont
	{LaForge}, \citenamefont {Cheong},\ and\ \citenamefont {Burch}}]{Zhao2011}%
\BibitemOpen
\bibfield  {author} {\bibinfo {author} {\bibfnamefont {S.~Y.~F.}\
		\bibnamefont {Zhao}}, \bibinfo {author} {\bibfnamefont {C.}~\bibnamefont
		{Beekman}}, \bibinfo {author} {\bibfnamefont {L.~J.}\ \bibnamefont
		{Sandilands}}, \bibinfo {author} {\bibfnamefont {J.~E.~J.}\ \bibnamefont
		{Bashucky}}, \bibinfo {author} {\bibfnamefont {D.}~\bibnamefont {Kwok}},
	\bibinfo {author} {\bibfnamefont {N.}~\bibnamefont {Lee}}, \bibinfo {author}
	{\bibfnamefont {A.~D.}\ \bibnamefont {LaForge}}, \bibinfo {author}
	{\bibfnamefont {S.~W.}\ \bibnamefont {Cheong}}, \ and\ \bibinfo {author}
	{\bibfnamefont {K.~S.}\ \bibnamefont {Burch}},\ }\bibfield  {title} {\enquote
	{\bibinfo {title} {Fabrication and characterization of topological insulator
			{Bi$_2$Se$_3$} nanocrystals},}\ }\href {\doibase 10.1063/1.3573868}
{\bibfield  {journal} {\bibinfo  {journal} {Applied Physics Letters}\
	}\textbf {\bibinfo {volume} {98}},\ \bibinfo {eid} {141911} (\bibinfo {year}
	{2011})}\BibitemShut {NoStop}%
\bibitem [{\citenamefont {Zhang}\ \emph {et~al.}(2011)\citenamefont {Zhang},
	\citenamefont {Peng}, \citenamefont {Soni}, \citenamefont {Zhao},
	\citenamefont {Xiong}, \citenamefont {Peng}, \citenamefont {Wang},
	\citenamefont {Dresselhaus},\ and\ \citenamefont {Xiong}}]{Zhang2011}%
\BibitemOpen
\bibfield  {author} {\bibinfo {author} {\bibfnamefont {Jun}\ \bibnamefont
		{Zhang}}, \bibinfo {author} {\bibfnamefont {Zeping}\ \bibnamefont {Peng}},
	\bibinfo {author} {\bibfnamefont {Ajay}\ \bibnamefont {Soni}}, \bibinfo
	{author} {\bibfnamefont {Yanyuan}\ \bibnamefont {Zhao}}, \bibinfo {author}
	{\bibfnamefont {Yi}~\bibnamefont {Xiong}}, \bibinfo {author} {\bibfnamefont
		{Bo}~\bibnamefont {Peng}}, \bibinfo {author} {\bibfnamefont {Jianbo}\
		\bibnamefont {Wang}}, \bibinfo {author} {\bibfnamefont {Mildred~S.}\
		\bibnamefont {Dresselhaus}}, \ and\ \bibinfo {author} {\bibfnamefont {Qihua}\
		\bibnamefont {Xiong}},\ }\bibfield  {title} {\enquote {\bibinfo {title}
		{{Raman} spectroscopy of few-quintuple layer topological insulator
			{Bi$_2$Se$_3$} nanoplatelets},}\ }\href {\doibase 10.1021/nl200773n}
{\bibfield  {journal} {\bibinfo  {journal} {Nano Letters}\ }\textbf {\bibinfo
		{volume} {11}},\ \bibinfo {pages} {2407--2414} (\bibinfo {year}
	{2011})}\BibitemShut {NoStop}%
\bibitem [{\citenamefont {Chis}\ \emph {et~al.}(2012)\citenamefont {Chis},
	\citenamefont {Sklyadneva}, \citenamefont {Kokh}, \citenamefont {Volodin},
	\citenamefont {Tereshchenko},\ and\ \citenamefont {Chulkov}}]{Chis2012}%
\BibitemOpen
\bibfield  {author} {\bibinfo {author} {\bibfnamefont {V.}~\bibnamefont
		{Chis}}, \bibinfo {author} {\bibfnamefont {I.~Yu.}\ \bibnamefont
		{Sklyadneva}}, \bibinfo {author} {\bibfnamefont {K.~A.}\ \bibnamefont
		{Kokh}}, \bibinfo {author} {\bibfnamefont {V.~A.}\ \bibnamefont {Volodin}},
	\bibinfo {author} {\bibfnamefont {O.~E.}\ \bibnamefont {Tereshchenko}}, \
	and\ \bibinfo {author} {\bibfnamefont {E.~V.}\ \bibnamefont {Chulkov}},\
}\bibfield  {title} {\enquote {\bibinfo {title} {Vibrations in binary and
		ternary topological insulators: First-principles calculations and {Raman}
		spectroscopy measurements},}\ }\href {\doibase 10.1103/PhysRevB.86.174304}
{\bibfield  {journal} {\bibinfo  {journal} {Phys. Rev. B}\ }\textbf {\bibinfo
		{volume} {86}},\ \bibinfo {pages} {174304} (\bibinfo {year}
	{2012})}\BibitemShut {NoStop}%
\bibitem [{\citenamefont {Huml\'{\i}\v{c}ek}\ \emph {et~al.}(2014)\citenamefont
	{Huml\'{\i}\v{c}ek}, \citenamefont {Hemzal}, \citenamefont {Dubroka},
	\citenamefont {Caha}, \citenamefont {Steiner}, \citenamefont {Bauer},\ and\
	\citenamefont {Springholz}}]{Humlicek2014}%
\BibitemOpen
\bibfield  {author} {\bibinfo {author} {\bibfnamefont {J}~\bibnamefont
		{Huml\'{\i}\v{c}ek}}, \bibinfo {author} {\bibfnamefont {D}~\bibnamefont
		{Hemzal}}, \bibinfo {author} {\bibfnamefont {A}~\bibnamefont {Dubroka}},
	\bibinfo {author} {\bibfnamefont {O}~\bibnamefont {Caha}}, \bibinfo {author}
	{\bibfnamefont {H}~\bibnamefont {Steiner}}, \bibinfo {author} {\bibfnamefont
		{G}~\bibnamefont {Bauer}}, \ and\ \bibinfo {author} {\bibfnamefont
		{G}~\bibnamefont {Springholz}},\ }\bibfield  {title} {\enquote {\bibinfo
		{title} {{Raman} and interband optical spectra of epitaxial layers of the
			topological insulators {Bi}$_2${Te}$_3$ and {Bi}$_2${Se}$_3$ on {BaF}$_2$
			substrates},}\ }\href {\doibase 10.1088/0031-8949/2014/T162/014007}
{\bibfield  {journal} {\bibinfo  {journal} {Physica Scripta}\ }\textbf
	{\bibinfo {volume} {2014}},\ \bibinfo {pages} {014007} (\bibinfo {year}
	{2014})}\BibitemShut {NoStop}%
\bibitem [{\citenamefont {Eddrief}\ \emph {et~al.}(2014)\citenamefont
	{Eddrief}, \citenamefont {Atkinson}, \citenamefont {Etgens},\ and\
	\citenamefont {Jusserand}}]{Eddrief2014}%
\BibitemOpen
\bibfield  {author} {\bibinfo {author} {\bibfnamefont {Mahmoud}\ \bibnamefont
		{Eddrief}}, \bibinfo {author} {\bibfnamefont {Paola}\ \bibnamefont
		{Atkinson}}, \bibinfo {author} {\bibfnamefont {Victor}\ \bibnamefont
		{Etgens}}, \ and\ \bibinfo {author} {\bibfnamefont {Bernard}\ \bibnamefont
		{Jusserand}},\ }\bibfield  {title} {\enquote {\bibinfo {title}
		{Low-temperature {R}aman fingerprints for few-quintuple layer topological
			insulator {Bi$_2$Se$_3$} films epitaxied on {GaAs}},}\ }\href {\doibase
	10.1088/0957-4484/25/24/245701} {\bibfield  {journal} {\bibinfo  {journal}
		{Nanotechnology}\ }\textbf {\bibinfo {volume} {25}},\ \bibinfo {pages}
	{245701} (\bibinfo {year} {2014})}\BibitemShut {NoStop}%
\bibitem [{\citenamefont {K\"{o}hler}\ and\ \citenamefont
	{Becker}(1974)}]{Kohler1974}%
\BibitemOpen
\bibfield  {author} {\bibinfo {author} {\bibfnamefont {H.}~\bibnamefont
		{K\"{o}hler}}\ and\ \bibinfo {author} {\bibfnamefont {C.~R.}\ \bibnamefont
		{Becker}},\ }\bibfield  {title} {\enquote {\bibinfo {title} {Optically active
			lattice vibrations in {Bi$_2$Se$_3$}},}\ }\href {\doibase
	10.1002/pssb.2220610218} {\bibfield  {journal} {\bibinfo  {journal} {physica
			status solidi (b)}\ }\textbf {\bibinfo {volume} {61}},\ \bibinfo {pages}
	{533--537} (\bibinfo {year} {1974})}\BibitemShut {NoStop}%
\bibitem [{\citenamefont {Richter}\ and\ \citenamefont
	{Becker}(1977)}]{Richter1977}%
\BibitemOpen
\bibfield  {author} {\bibinfo {author} {\bibfnamefont {W.}~\bibnamefont
		{Richter}}\ and\ \bibinfo {author} {\bibfnamefont {C.~R.}\ \bibnamefont
		{Becker}},\ }\bibfield  {title} {\enquote {\bibinfo {title} {{A Raman and
				far-infrared investigation of phonons in the rhombohedral V$_2$–VI$_3$
				compounds Bi$_2$Te$_3$, Bi$_2$Se$_3$, Sb$_2$Te$_3$ and
				Bi$_2$(Te$_{1-x}$Se$_x$)$_3$ ($0 < x < 1$), (Bi$_{1-y}$Sb$_y$)$_2$Te$_3$ ($0
				< y < 1$)}},}\ }\href {\doibase 10.1002/pssb.2220840226} {\bibfield
	{journal} {\bibinfo  {journal} {physica status solidi (b)}\ }\textbf
	{\bibinfo {volume} {84}},\ \bibinfo {pages} {619--628} (\bibinfo {year}
	{1977})}\BibitemShut {NoStop}%
\bibitem [{\citenamefont {LaForge}\ \emph {et~al.}(2010)\citenamefont
	{LaForge}, \citenamefont {Frenzel}, \citenamefont {Pursley}, \citenamefont
	{Lin}, \citenamefont {Liu}, \citenamefont {Shi},\ and\ \citenamefont
	{Basov}}]{LaForge2010}%
\BibitemOpen
\bibfield  {author} {\bibinfo {author} {\bibfnamefont {A.~D.}\ \bibnamefont
		{LaForge}}, \bibinfo {author} {\bibfnamefont {A.}~\bibnamefont {Frenzel}},
	\bibinfo {author} {\bibfnamefont {B.~C.}\ \bibnamefont {Pursley}}, \bibinfo
	{author} {\bibfnamefont {Tao}\ \bibnamefont {Lin}}, \bibinfo {author}
	{\bibfnamefont {Xinfei}\ \bibnamefont {Liu}}, \bibinfo {author}
	{\bibfnamefont {Jing}\ \bibnamefont {Shi}}, \ and\ \bibinfo {author}
	{\bibfnamefont {D.~N.}\ \bibnamefont {Basov}},\ }\bibfield  {title} {\enquote
	{\bibinfo {title} {Optical characterization of {Bi}$_2${Se}$_3$ in a magnetic
			field: Infrared evidence for magnetoelectric coupling in a topological
			insulator material},}\ }\href {\doibase 10.1103/PhysRevB.81.125120}
{\bibfield  {journal} {\bibinfo  {journal} {Phys. Rev. B}\ }\textbf {\bibinfo
		{volume} {81}},\ \bibinfo {pages} {125120} (\bibinfo {year}
	{2010})}\BibitemShut {NoStop}%
\bibitem [{\citenamefont {Gnezdilov}\ \emph {et~al.}(2011)\citenamefont
	{Gnezdilov}, \citenamefont {Pashkevich}, \citenamefont {Berger},
	\citenamefont {Pomjakushina}, \citenamefont {Conder},\ and\ \citenamefont
	{Lemmens}}]{Gnezdilov2011}%
\BibitemOpen
\bibfield  {author} {\bibinfo {author} {\bibfnamefont {V.}~\bibnamefont
		{Gnezdilov}}, \bibinfo {author} {\bibfnamefont {Yu.~G.}\ \bibnamefont
		{Pashkevich}}, \bibinfo {author} {\bibfnamefont {H.}~\bibnamefont {Berger}},
	\bibinfo {author} {\bibfnamefont {E.}~\bibnamefont {Pomjakushina}}, \bibinfo
	{author} {\bibfnamefont {K.}~\bibnamefont {Conder}}, \ and\ \bibinfo {author}
	{\bibfnamefont {P.}~\bibnamefont {Lemmens}},\ }\bibfield  {title} {\enquote
	{\bibinfo {title} {Helical fluctuations in the {Raman} response of the
			topological insulator {Bi$_{2}$Se$_{3}$}},}\ }\href {\doibase
	10.1103/PhysRevB.84.195118} {\bibfield  {journal} {\bibinfo  {journal} {Phys.
			Rev. B}\ }\textbf {\bibinfo {volume} {84}},\ \bibinfo {pages} {195118}
	(\bibinfo {year} {2011})}\BibitemShut {NoStop}%
\bibitem [{\citenamefont {Kim}\ \emph {et~al.}(2012)\citenamefont {Kim},
	\citenamefont {Chen}, \citenamefont {Wang}, \citenamefont {Shi},
	\citenamefont {Miotkowski}, \citenamefont {Chen}, \citenamefont {Sharma},
	\citenamefont {Sharma}, \citenamefont {Hekmaty}, \citenamefont {Jiang},\ and\
	\citenamefont {Smirnov}}]{Kim2012}%
\BibitemOpen
\bibfield  {author} {\bibinfo {author} {\bibfnamefont {Y.}~\bibnamefont
		{Kim}}, \bibinfo {author} {\bibfnamefont {X.}~\bibnamefont {Chen}}, \bibinfo
	{author} {\bibfnamefont {Z.}~\bibnamefont {Wang}}, \bibinfo {author}
	{\bibfnamefont {J.}~\bibnamefont {Shi}}, \bibinfo {author} {\bibfnamefont
		{I.}~\bibnamefont {Miotkowski}}, \bibinfo {author} {\bibfnamefont {Y.~P.}\
		\bibnamefont {Chen}}, \bibinfo {author} {\bibfnamefont {P.~A.}\ \bibnamefont
		{Sharma}}, \bibinfo {author} {\bibfnamefont {A.~L.~Lima}\ \bibnamefont
		{Sharma}}, \bibinfo {author} {\bibfnamefont {M.~A.}\ \bibnamefont {Hekmaty}},
	\bibinfo {author} {\bibfnamefont {Z.}~\bibnamefont {Jiang}}, \ and\ \bibinfo
	{author} {\bibfnamefont {D.}~\bibnamefont {Smirnov}},\ }\bibfield  {title}
{\enquote {\bibinfo {title} {Temperature dependence of {Raman}-active optical
			phonons in {Bi}$_2${Se}$_3$ and {Sb}$_2${Te}$_3$},}\ }\href {\doibase
	10.1063/1.3685465} {\bibfield  {journal} {\bibinfo  {journal} {Applied
			Physics Letters}\ }\textbf {\bibinfo {volume} {100}},\ \bibinfo {pages}
	{071907} (\bibinfo {year} {2012})}\BibitemShut {NoStop}%
\bibitem [{\citenamefont {Irfan}\ \emph {et~al.}(2014)\citenamefont {Irfan},
	\citenamefont {Sahoo}, \citenamefont {Gaur}, \citenamefont {Ahmadi},
	\citenamefont {Guinel}, \citenamefont {Katiyar},\ and\ \citenamefont
	{Chatterjee}}]{Irfan2014}%
\BibitemOpen
\bibfield  {author} {\bibinfo {author} {\bibfnamefont {Bushra}\ \bibnamefont
		{Irfan}}, \bibinfo {author} {\bibfnamefont {Satyaprakash}\ \bibnamefont
		{Sahoo}}, \bibinfo {author} {\bibfnamefont {Anand P.~S.}\ \bibnamefont
		{Gaur}}, \bibinfo {author} {\bibfnamefont {Majid}\ \bibnamefont {Ahmadi}},
	\bibinfo {author} {\bibfnamefont {Maxime J.-F.}\ \bibnamefont {Guinel}},
	\bibinfo {author} {\bibfnamefont {Ram~S.}\ \bibnamefont {Katiyar}}, \ and\
	\bibinfo {author} {\bibfnamefont {Ratnamala}\ \bibnamefont {Chatterjee}},\
}\bibfield  {title} {\enquote {\bibinfo {title} {Temperature dependent
		{Raman} scattering studies of three dimensional topological insulators
		{Bi}$_2${Se}$_3$},}\ }\href {\doibase 10.1063/1.4871860} {\bibfield
{journal} {\bibinfo  {journal} {Journal of Applied Physics}\ }\textbf
{\bibinfo {volume} {115}},\ \bibinfo {pages} {173506} (\bibinfo {year}
{2014})}\BibitemShut {NoStop}%
\bibitem [{\citenamefont {Yan}\ \emph {et~al.}(2015)\citenamefont {Yan},
	\citenamefont {Zhou}, \citenamefont {Jin}, \citenamefont {Li}, \citenamefont
	{Ke}, \citenamefont {Van~Tendeloo}, \citenamefont {Liu}, \citenamefont {Yu},
	\citenamefont {Dressel},\ and\ \citenamefont {Liao}}]{Yan2015}%
\BibitemOpen
\bibfield  {author} {\bibinfo {author} {\bibfnamefont {Yuan}\ \bibnamefont
		{Yan}}, \bibinfo {author} {\bibfnamefont {Xu}~\bibnamefont {Zhou}}, \bibinfo
	{author} {\bibfnamefont {Han}\ \bibnamefont {Jin}}, \bibinfo {author}
	{\bibfnamefont {Cai-Zhen}\ \bibnamefont {Li}}, \bibinfo {author}
	{\bibfnamefont {Xiaoxing}\ \bibnamefont {Ke}}, \bibinfo {author}
	{\bibfnamefont {Gustaaf}\ \bibnamefont {Van~Tendeloo}}, \bibinfo {author}
	{\bibfnamefont {Kaihui}\ \bibnamefont {Liu}}, \bibinfo {author}
	{\bibfnamefont {Dapeng}\ \bibnamefont {Yu}}, \bibinfo {author} {\bibfnamefont
		{Martin}\ \bibnamefont {Dressel}}, \ and\ \bibinfo {author} {\bibfnamefont
		{Zhi-Min}\ \bibnamefont {Liao}},\ }\bibfield  {title} {\enquote {\bibinfo
		{title} {Surface-facet-dependent phonon deformation potential in individual
			strained topological insulator {Bi}$_2${Se}$_3$ nanoribbons},}\ }\href
{\doibase 10.1021/acsnano.5b04057} {\bibfield  {journal} {\bibinfo  {journal}
		{ACS Nano}\ }\textbf {\bibinfo {volume} {9}},\ \bibinfo {pages}
	{10244--10251} (\bibinfo {year} {2015})}\BibitemShut {NoStop}%
\bibitem [{\citenamefont {Zhang}\ \emph {et~al.}(2016)\citenamefont {Zhang},
	\citenamefont {Tan}, \citenamefont {Wu}, \citenamefont {Shi},\ and\
	\citenamefont {Tan}}]{Zhang2016}%
\BibitemOpen
\bibfield  {author} {\bibinfo {author} {\bibfnamefont {Xin}\ \bibnamefont
		{Zhang}}, \bibinfo {author} {\bibfnamefont {Qing-Hai}\ \bibnamefont {Tan}},
	\bibinfo {author} {\bibfnamefont {Jiang-Bin}\ \bibnamefont {Wu}}, \bibinfo
	{author} {\bibfnamefont {Wei}\ \bibnamefont {Shi}}, \ and\ \bibinfo {author}
	{\bibfnamefont {Ping-Heng}\ \bibnamefont {Tan}},\ }\bibfield  {title}
{\enquote {\bibinfo {title} {Review on the {Raman} spectroscopy of different
			types of layered materials},}\ }\href {\doibase 10.1039/C5NR07205K}
{\bibfield  {journal} {\bibinfo  {journal} {Nanoscale}\ }\textbf {\bibinfo
		{volume} {8}},\ \bibinfo {pages} {6435--6450} (\bibinfo {year}
	{2016})}\BibitemShut {NoStop}%
\bibitem [{\citenamefont {Zhu}\ \emph {et~al.}(2011)\citenamefont {Zhu},
	\citenamefont {Santos}, \citenamefont {Sankar}, \citenamefont {Chikara},
	\citenamefont {Howard}, \citenamefont {Chou}, \citenamefont {Chamon},\ and\
	\citenamefont {El-Batanouny}}]{Zhu2011}%
\BibitemOpen
\bibfield  {author} {\bibinfo {author} {\bibfnamefont {Xuetao}\ \bibnamefont
		{Zhu}}, \bibinfo {author} {\bibfnamefont {L.}~\bibnamefont {Santos}},
	\bibinfo {author} {\bibfnamefont {R.}~\bibnamefont {Sankar}}, \bibinfo
	{author} {\bibfnamefont {S.}~\bibnamefont {Chikara}}, \bibinfo {author}
	{\bibfnamefont {C.~.}\ \bibnamefont {Howard}}, \bibinfo {author}
	{\bibfnamefont {F.~C.}\ \bibnamefont {Chou}}, \bibinfo {author}
	{\bibfnamefont {C.}~\bibnamefont {Chamon}}, \ and\ \bibinfo {author}
	{\bibfnamefont {M.}~\bibnamefont {El-Batanouny}},\ }\bibfield  {title}
{\enquote {\bibinfo {title} {Interaction of phonons and {D}irac fermions on
			the surface of {Bi}$_2${Se}$_3$: A strong {K}ohn anomaly},}\ }\href {\doibase
	10.1103/PhysRevLett.107.186102} {\bibfield  {journal} {\bibinfo  {journal}
		{Phys. Rev. Lett.}\ }\textbf {\bibinfo {volume} {107}},\ \bibinfo {pages}
	{186102} (\bibinfo {year} {2011})}\BibitemShut {NoStop}%
\bibitem [{\citenamefont {Zhu}\ \emph {et~al.}(2012)\citenamefont {Zhu},
	\citenamefont {Santos}, \citenamefont {Howard}, \citenamefont {Sankar},
	\citenamefont {Chou}, \citenamefont {Chamon},\ and\ \citenamefont
	{El-Batanouny}}]{Zhu2012}%
\BibitemOpen
\bibfield  {author} {\bibinfo {author} {\bibfnamefont {Xuetao}\ \bibnamefont
		{Zhu}}, \bibinfo {author} {\bibfnamefont {L.}~\bibnamefont {Santos}},
	\bibinfo {author} {\bibfnamefont {C.}~\bibnamefont {Howard}}, \bibinfo
	{author} {\bibfnamefont {R.}~\bibnamefont {Sankar}}, \bibinfo {author}
	{\bibfnamefont {F.~C.}\ \bibnamefont {Chou}}, \bibinfo {author}
	{\bibfnamefont {C.}~\bibnamefont {Chamon}}, \ and\ \bibinfo {author}
	{\bibfnamefont {M.}~\bibnamefont {El-Batanouny}},\ }\bibfield  {title}
{\enquote {\bibinfo {title} {Electron-phonon coupling on the surface of the
			topological insulator {Bi}$_2${Se}$_3$ determined from surface-phonon
			dispersion measurements},}\ }\href {\doibase 10.1103/PhysRevLett.108.185501}
{\bibfield  {journal} {\bibinfo  {journal} {Phys. Rev. Lett.}\ }\textbf
	{\bibinfo {volume} {108}},\ \bibinfo {pages} {185501} (\bibinfo {year}
	{2012})}\BibitemShut {NoStop}%
\bibitem [{\citenamefont {Howard}\ \emph {et~al.}(2013)\citenamefont {Howard},
	\citenamefont {El-Batanouny}, \citenamefont {Sankar},\ and\ \citenamefont
	{Chou}}]{Howard2013}%
\BibitemOpen
\bibfield  {author} {\bibinfo {author} {\bibfnamefont {C.}~\bibnamefont
		{Howard}}, \bibinfo {author} {\bibfnamefont {M.}~\bibnamefont
		{El-Batanouny}}, \bibinfo {author} {\bibfnamefont {R.}~\bibnamefont
		{Sankar}}, \ and\ \bibinfo {author} {\bibfnamefont {F.~C.}\ \bibnamefont
		{Chou}},\ }\bibfield  {title} {\enquote {\bibinfo {title} {Anomalous behavior
			in the phonon dispersion of the (001) surface of {Bi}$_{2}${Te}$_{3}$
			determined from helium atom-surface scattering measurements},}\ }\href
{\doibase 10.1103/PhysRevB.88.035402} {\bibfield  {journal} {\bibinfo
		{journal} {Phys. Rev. B}\ }\textbf {\bibinfo {volume} {88}},\ \bibinfo
	{pages} {035402} (\bibinfo {year} {2013})}\BibitemShut {NoStop}%
\bibitem [{\citenamefont {Hatch}\ \emph {et~al.}(2011)\citenamefont {Hatch},
	\citenamefont {Bianchi}, \citenamefont {Guan}, \citenamefont {Bao},
	\citenamefont {Mi}, \citenamefont {Iversen}, \citenamefont {Nilsson},
	\citenamefont {Hornek\ae{}r},\ and\ \citenamefont {Hofmann}}]{Hatch2011}%
\BibitemOpen
\bibfield  {author} {\bibinfo {author} {\bibfnamefont {Richard~C.}\
		\bibnamefont {Hatch}}, \bibinfo {author} {\bibfnamefont {Marco}\ \bibnamefont
		{Bianchi}}, \bibinfo {author} {\bibfnamefont {Dandan}\ \bibnamefont {Guan}},
	\bibinfo {author} {\bibfnamefont {Shining}\ \bibnamefont {Bao}}, \bibinfo
	{author} {\bibfnamefont {Jianli}\ \bibnamefont {Mi}}, \bibinfo {author}
	{\bibfnamefont {Bo~Brummerstedt}\ \bibnamefont {Iversen}}, \bibinfo {author}
	{\bibfnamefont {Louis}\ \bibnamefont {Nilsson}}, \bibinfo {author}
	{\bibfnamefont {Liv}\ \bibnamefont {Hornek\ae{}r}}, \ and\ \bibinfo {author}
	{\bibfnamefont {Philip}\ \bibnamefont {Hofmann}},\ }\bibfield  {title}
{\enquote {\bibinfo {title} {Stability of the {Bi}$_2${Se}$_3$(111)
			topological state: Electron-phonon and electron-defect scattering},}\ }\href
{\doibase 10.1103/PhysRevB.83.241303} {\bibfield  {journal} {\bibinfo
		{journal} {Phys. Rev. B}\ }\textbf {\bibinfo {volume} {83}},\ \bibinfo
	{pages} {241303} (\bibinfo {year} {2011})}\BibitemShut {NoStop}%
\bibitem [{\citenamefont {Sobota}\ \emph {et~al.}(2014)\citenamefont {Sobota},
	\citenamefont {Yang}, \citenamefont {Leuenberger}, \citenamefont {Kemper},
	\citenamefont {Analytis}, \citenamefont {Fisher}, \citenamefont {Kirchmann},
	\citenamefont {Devereaux},\ and\ \citenamefont {Shen}}]{Sobota2014}%
\BibitemOpen
\bibfield  {author} {\bibinfo {author} {\bibfnamefont {J.~A.}\ \bibnamefont
		{Sobota}}, \bibinfo {author} {\bibfnamefont {S.-L.}\ \bibnamefont {Yang}},
	\bibinfo {author} {\bibfnamefont {D.}~\bibnamefont {Leuenberger}}, \bibinfo
	{author} {\bibfnamefont {A.~F.}\ \bibnamefont {Kemper}}, \bibinfo {author}
	{\bibfnamefont {J.~G.}\ \bibnamefont {Analytis}}, \bibinfo {author}
	{\bibfnamefont {I.~R.}\ \bibnamefont {Fisher}}, \bibinfo {author}
	{\bibfnamefont {P.~S.}\ \bibnamefont {Kirchmann}}, \bibinfo {author}
	{\bibfnamefont {T.~P.}\ \bibnamefont {Devereaux}}, \ and\ \bibinfo {author}
	{\bibfnamefont {Z.-X.}\ \bibnamefont {Shen}},\ }\bibfield  {title} {\enquote
	{\bibinfo {title} {Distinguishing bulk and surface electron-phonon coupling
			in the topological insulator {Bi}$_2${Se}$_3$ using time-resolved
			photoemission spectroscopy},}\ }\href {\doibase
	10.1103/PhysRevLett.113.157401} {\bibfield  {journal} {\bibinfo  {journal}
		{Phys. Rev. Lett.}\ }\textbf {\bibinfo {volume} {113}},\ \bibinfo {pages}
	{157401} (\bibinfo {year} {2014})}\BibitemShut {NoStop}%
\bibitem [{\citenamefont {Chen}\ \emph {et~al.}(2013)\citenamefont {Chen},
	\citenamefont {Xie}, \citenamefont {Feng}, \citenamefont {Yi}, \citenamefont
	{Liang}, \citenamefont {He}, \citenamefont {Mou}, \citenamefont {He},
	\citenamefont {Peng}, \citenamefont {Liu}, \citenamefont {Liu}, \citenamefont
	{Zhao}, \citenamefont {Liu}, \citenamefont {Dong}, \citenamefont {Zhang},
	\citenamefont {Yu}, \citenamefont {Wang}, \citenamefont {Peng}, \citenamefont
	{Wang}, \citenamefont {Zhang}, \citenamefont {Yang}, \citenamefont {Chen},
	\citenamefont {Xu},\ and\ \citenamefont {Zhou}}]{Chen2013}%
\BibitemOpen
\bibfield  {author} {\bibinfo {author} {\bibfnamefont {Chaoyu}\ \bibnamefont
		{Chen}}, \bibinfo {author} {\bibfnamefont {Zhuojin}\ \bibnamefont {Xie}},
	\bibinfo {author} {\bibfnamefont {Ya}~\bibnamefont {Feng}}, \bibinfo {author}
	{\bibfnamefont {Hemian}\ \bibnamefont {Yi}}, \bibinfo {author} {\bibfnamefont
		{Aiji}\ \bibnamefont {Liang}}, \bibinfo {author} {\bibfnamefont {Shaolong}\
		\bibnamefont {He}}, \bibinfo {author} {\bibfnamefont {Daixiang}\ \bibnamefont
		{Mou}}, \bibinfo {author} {\bibfnamefont {Junfeng}\ \bibnamefont {He}},
	\bibinfo {author} {\bibfnamefont {Yingying}\ \bibnamefont {Peng}}, \bibinfo
	{author} {\bibfnamefont {Xu}~\bibnamefont {Liu}}, \bibinfo {author}
	{\bibfnamefont {Yan}\ \bibnamefont {Liu}}, \bibinfo {author} {\bibfnamefont
		{Lin}\ \bibnamefont {Zhao}}, \bibinfo {author} {\bibfnamefont {Guodong}\
		\bibnamefont {Liu}}, \bibinfo {author} {\bibfnamefont {Xiaoli}\ \bibnamefont
		{Dong}}, \bibinfo {author} {\bibfnamefont {Jun}\ \bibnamefont {Zhang}},
	\bibinfo {author} {\bibfnamefont {Li}~\bibnamefont {Yu}}, \bibinfo {author}
	{\bibfnamefont {Xiaoyang}\ \bibnamefont {Wang}}, \bibinfo {author}
	{\bibfnamefont {Qinjun}\ \bibnamefont {Peng}}, \bibinfo {author}
	{\bibfnamefont {Zhimin}\ \bibnamefont {Wang}}, \bibinfo {author}
	{\bibfnamefont {Shenjin}\ \bibnamefont {Zhang}}, \bibinfo {author}
	{\bibfnamefont {Feng}\ \bibnamefont {Yang}}, \bibinfo {author} {\bibfnamefont
		{Chuangtian}\ \bibnamefont {Chen}}, \bibinfo {author} {\bibfnamefont {Zuyan}\
		\bibnamefont {Xu}}, \ and\ \bibinfo {author} {\bibfnamefont {X.~J.}\
		\bibnamefont {Zhou}},\ }\bibfield  {title} {\enquote {\bibinfo {title}
		{Tunable {D}irac fermion dynamics in topological insulators},}\ }\href
{http://dx.doi.org/10.1038/srep02411} {\bibfield  {journal} {\bibinfo
		{journal} {Scientific Reports}\ }\textbf {\bibinfo {volume} {3}},\ \bibinfo
	{pages} {2411} (\bibinfo {year} {2013})}\BibitemShut {NoStop}%
\bibitem [{\citenamefont {Kondo}\ \emph {et~al.}(2013)\citenamefont {Kondo},
	\citenamefont {Nakashima}, \citenamefont {Ota}, \citenamefont {Ishida},
	\citenamefont {Malaeb}, \citenamefont {Okazaki}, \citenamefont {Shin},
	\citenamefont {Kriener}, \citenamefont {Sasaki}, \citenamefont {Segawa},\
	and\ \citenamefont {Ando}}]{Kondo2013}%
\BibitemOpen
\bibfield  {author} {\bibinfo {author} {\bibfnamefont {Takeshi}\ \bibnamefont
		{Kondo}}, \bibinfo {author} {\bibfnamefont {Y.}~\bibnamefont {Nakashima}},
	\bibinfo {author} {\bibfnamefont {Y.}~\bibnamefont {Ota}}, \bibinfo {author}
	{\bibfnamefont {Y.}~\bibnamefont {Ishida}}, \bibinfo {author} {\bibfnamefont
		{W.}~\bibnamefont {Malaeb}}, \bibinfo {author} {\bibfnamefont
		{K.}~\bibnamefont {Okazaki}}, \bibinfo {author} {\bibfnamefont
		{S.}~\bibnamefont {Shin}}, \bibinfo {author} {\bibfnamefont {M.}~\bibnamefont
		{Kriener}}, \bibinfo {author} {\bibfnamefont {Satoshi}\ \bibnamefont
		{Sasaki}}, \bibinfo {author} {\bibfnamefont {Kouji}\ \bibnamefont {Segawa}},
	\ and\ \bibinfo {author} {\bibfnamefont {Yoichi}\ \bibnamefont {Ando}},\
}\bibfield  {title} {\enquote {\bibinfo {title} {Anomalous dressing of
		{D}irac fermions in the topological surface state of {Bi}$_2${Se}$_3$,
		{Bi}$_2${Te}$_3$, and {Cu}-doped {Bi}$_2${Se}$_3$},}\ }\href {\doibase
10.1103/PhysRevLett.110.217601} {\bibfield  {journal} {\bibinfo  {journal}
	{Phys. Rev. Lett.}\ }\textbf {\bibinfo {volume} {110}},\ \bibinfo {pages}
{217601} (\bibinfo {year} {2013})}\BibitemShut {NoStop}%
\bibitem [{\citenamefont {Kogar}\ \emph {et~al.}(2015)\citenamefont {Kogar},
	\citenamefont {Vig}, \citenamefont {Thaler}, \citenamefont {Wong},
	\citenamefont {Xiao}, \citenamefont {{Reig-i-Plessis}}, \citenamefont {Cho},
	\citenamefont {Valla}, \citenamefont {Pan}, \citenamefont {Schneeloch},
	\citenamefont {Zhong}, \citenamefont {Gu}, \citenamefont {Hughes},
	\citenamefont {MacDougall}, \citenamefont {Chiang},\ and\ \citenamefont
	{Abbamonte}}]{Kogar2015}%
\BibitemOpen
\bibfield  {author} {\bibinfo {author} {\bibfnamefont {A.}~\bibnamefont
		{Kogar}}, \bibinfo {author} {\bibfnamefont {S.}~\bibnamefont {Vig}}, \bibinfo
	{author} {\bibfnamefont {A.}~\bibnamefont {Thaler}}, \bibinfo {author}
	{\bibfnamefont {M.~H.}\ \bibnamefont {Wong}}, \bibinfo {author}
	{\bibfnamefont {Y.}~\bibnamefont {Xiao}}, \bibinfo {author} {\bibfnamefont
		{D.}~\bibnamefont {{Reig-i-Plessis}}}, \bibinfo {author} {\bibfnamefont
		{G.~Y.}\ \bibnamefont {Cho}}, \bibinfo {author} {\bibfnamefont
		{T.}~\bibnamefont {Valla}}, \bibinfo {author} {\bibfnamefont
		{Z.}~\bibnamefont {Pan}}, \bibinfo {author} {\bibfnamefont {J.}~\bibnamefont
		{Schneeloch}}, \bibinfo {author} {\bibfnamefont {R.}~\bibnamefont {Zhong}},
	\bibinfo {author} {\bibfnamefont {G.~D.}\ \bibnamefont {Gu}}, \bibinfo
	{author} {\bibfnamefont {T.~L.}\ \bibnamefont {Hughes}}, \bibinfo {author}
	{\bibfnamefont {G.~J.}\ \bibnamefont {MacDougall}}, \bibinfo {author}
	{\bibfnamefont {T.-C.}\ \bibnamefont {Chiang}}, \ and\ \bibinfo {author}
	{\bibfnamefont {P.}~\bibnamefont {Abbamonte}},\ }\bibfield  {title} {\enquote
	{\bibinfo {title} {Surface collective modes in the topological insulators
			{Bi}$_2${Se}$_3$ and {Bi}$_{0.5}${Sb}$_{1.5}${Te}$_{3-x}${Se}$_{x}$},}\
}\href {\doibase 10.1103/PhysRevLett.115.257402} {\bibfield  {journal}
{\bibinfo  {journal} {Phys. Rev. Lett.}\ }\textbf {\bibinfo {volume} {115}},\
\bibinfo {pages} {257402} (\bibinfo {year} {2015})}\BibitemShut {NoStop}%
\bibitem [{\citenamefont {Esser}\ and\ \citenamefont
	{Richter}(1999)}]{Esser1999}%
\BibitemOpen
\bibfield  {author} {\bibinfo {author} {\bibfnamefont {Norbert}\ \bibnamefont
		{Esser}}\ and\ \bibinfo {author} {\bibfnamefont {Wolfgang}\ \bibnamefont
		{Richter}},\ }\bibfield  {title} {\enquote {\bibinfo {title} {Raman
			scattering from surface phonons},}\ }in\ \href@noop {} {\emph {\bibinfo
		{booktitle} {Light scattering in solids VIII}}},\ \bibinfo {editor} {edited
	by\ \bibinfo {editor} {\bibfnamefont {Manuel}\ \bibnamefont {Cardona}}\ and\
	\bibinfo {editor} {\bibfnamefont {Gernot}\ \bibnamefont {G\"{u}ntherodt}}}\
(\bibinfo  {publisher} {Springer-Verlag, Berlin},\ \bibinfo {year} {1999})\
pp.\ \bibinfo {pages} {96--168}\BibitemShut {NoStop}%
\bibitem [{\citenamefont {Liebhaber}\ \emph {et~al.}(2014)\citenamefont
	{Liebhaber}, \citenamefont {Bass}, \citenamefont {Bayersdorfer},
	\citenamefont {Geurts}, \citenamefont {Speiser}, \citenamefont {R\"athel},
	\citenamefont {Baumann}, \citenamefont {Chandola},\ and\ \citenamefont
	{Esser}}]{Liebhaber2014}%
\BibitemOpen
\bibfield  {author} {\bibinfo {author} {\bibfnamefont {M.}~\bibnamefont
		{Liebhaber}}, \bibinfo {author} {\bibfnamefont {U.}~\bibnamefont {Bass}},
	\bibinfo {author} {\bibfnamefont {P.}~\bibnamefont {Bayersdorfer}}, \bibinfo
	{author} {\bibfnamefont {J.}~\bibnamefont {Geurts}}, \bibinfo {author}
	{\bibfnamefont {E.}~\bibnamefont {Speiser}}, \bibinfo {author} {\bibfnamefont
		{J.}~\bibnamefont {R\"athel}}, \bibinfo {author} {\bibfnamefont
		{A.}~\bibnamefont {Baumann}}, \bibinfo {author} {\bibfnamefont
		{S.}~\bibnamefont {Chandola}}, \ and\ \bibinfo {author} {\bibfnamefont
		{N.}~\bibnamefont {Esser}},\ }\bibfield  {title} {\enquote {\bibinfo {title}
		{Surface phonons of the {Si}(111)-($7\times 7$) reconstruction observed by
			{Raman} spectroscopy},}\ }\href {\doibase 10.1103/PhysRevB.89.045313}
{\bibfield  {journal} {\bibinfo  {journal} {Phys. Rev. B}\ }\textbf {\bibinfo
		{volume} {89}},\ \bibinfo {pages} {045313} (\bibinfo {year}
	{2014})}\BibitemShut {NoStop}%
\bibitem [{\citenamefont {Lo{\v{s}}{\v{T}}{\'a}k}\ \emph
	{et~al.}(1990)\citenamefont {Lo{\v{s}}{\v{T}}{\'a}k}, \citenamefont
	{Bene{\v{s}}}, \citenamefont {Civi{\v{s}}},\ and\ \citenamefont
	{S{\"u}ssmann}}]{Lostak1990}%
\BibitemOpen
\bibfield  {author} {\bibinfo {author} {\bibfnamefont {P.}~\bibnamefont
		{Lo{\v{s}}{\v{T}}{\'a}k}}, \bibinfo {author} {\bibfnamefont {L.}~\bibnamefont
		{Bene{\v{s}}}}, \bibinfo {author} {\bibfnamefont {S.}~\bibnamefont
		{Civi{\v{s}}}}, \ and\ \bibinfo {author} {\bibfnamefont {H.}~\bibnamefont
		{S{\"u}ssmann}},\ }\bibfield  {title} {\enquote {\bibinfo {title}
		{Preparation and some physical properties of {Bi}$_{2−x}${In}$_x${Se}$_3$
			single crystals},}\ }\href {\doibase 10.1007/BF00544220} {\bibfield
	{journal} {\bibinfo  {journal} {Journal of Materials Science}\ }\textbf
	{\bibinfo {volume} {25}},\ \bibinfo {pages} {277--282} (\bibinfo {year}
	{1990})}\BibitemShut {NoStop}%
\bibitem [{\citenamefont {Dai}\ \emph {et~al.}(2016)\citenamefont {Dai},
	\citenamefont {West}, \citenamefont {Wang}, \citenamefont {Wang},
	\citenamefont {Kwok}, \citenamefont {Cheong}, \citenamefont {Zhang},\ and\
	\citenamefont {Wu}}]{Dai2016}%
\BibitemOpen
\bibfield  {author} {\bibinfo {author} {\bibfnamefont {Jixia}\ \bibnamefont
		{Dai}}, \bibinfo {author} {\bibfnamefont {Damien}\ \bibnamefont {West}},
	\bibinfo {author} {\bibfnamefont {Xueyun}\ \bibnamefont {Wang}}, \bibinfo
	{author} {\bibfnamefont {Yazhong}\ \bibnamefont {Wang}}, \bibinfo {author}
	{\bibfnamefont {Daniel}\ \bibnamefont {Kwok}}, \bibinfo {author}
	{\bibfnamefont {S.-W.}\ \bibnamefont {Cheong}}, \bibinfo {author}
	{\bibfnamefont {S.~B.}\ \bibnamefont {Zhang}}, \ and\ \bibinfo {author}
	{\bibfnamefont {Weida}\ \bibnamefont {Wu}},\ }\bibfield  {title} {\enquote
	{\bibinfo {title} {Toward the intrinsic limit of the topological insulator
			{Bi}$_2${Se}$_3$},}\ }\href {\doibase 10.1103/PhysRevLett.117.106401}
{\bibfield  {journal} {\bibinfo  {journal} {Phys. Rev. Lett.}\ }\textbf
	{\bibinfo {volume} {117}},\ \bibinfo {pages} {106401} (\bibinfo {year}
	{2016})}\BibitemShut {NoStop}%
\bibitem [{\citenamefont {Brahlek}\ \emph {et~al.}(2012)\citenamefont
	{Brahlek}, \citenamefont {Bansal}, \citenamefont {Koirala}, \citenamefont
	{Xu}, \citenamefont {Neupane}, \citenamefont {Liu}, \citenamefont {Hasan},\
	and\ \citenamefont {Oh}}]{Brahlek2012}%
\BibitemOpen
\bibfield  {author} {\bibinfo {author} {\bibfnamefont {Matthew}\ \bibnamefont
		{Brahlek}}, \bibinfo {author} {\bibfnamefont {Namrata}\ \bibnamefont
		{Bansal}}, \bibinfo {author} {\bibfnamefont {Nikesh}\ \bibnamefont
		{Koirala}}, \bibinfo {author} {\bibfnamefont {Su-Yang}\ \bibnamefont {Xu}},
	\bibinfo {author} {\bibfnamefont {Madhab}\ \bibnamefont {Neupane}}, \bibinfo
	{author} {\bibfnamefont {Chang}\ \bibnamefont {Liu}}, \bibinfo {author}
	{\bibfnamefont {{M. Zahid}}\ \bibnamefont {Hasan}}, \ and\ \bibinfo {author}
	{\bibfnamefont {Seongshik}\ \bibnamefont {Oh}},\ }\bibfield  {title}
{\enquote {\bibinfo {title} {Topological-metal to band-insulator transition
			in ({Bi}$_{1-x}${In}$_x$)$_2${Se}$_3$ thin films},}\ }\href {\doibase
	10.1103/PhysRevLett.109.186403} {\bibfield  {journal} {\bibinfo  {journal}
		{Phys. Rev. Lett.}\ }\textbf {\bibinfo {volume} {109}},\ \bibinfo {pages}
	{186403} (\bibinfo {year} {2012})}\BibitemShut {NoStop}%
\bibitem [{\citenamefont {Bansal}\ \emph {et~al.}(2012)\citenamefont {Bansal},
	\citenamefont {Kim}, \citenamefont {Brahlek}, \citenamefont {Edrey},\ and\
	\citenamefont {Oh}}]{Bansal2012}%
\BibitemOpen
\bibfield  {author} {\bibinfo {author} {\bibfnamefont {Namrata}\ \bibnamefont
		{Bansal}}, \bibinfo {author} {\bibfnamefont {Yong~Seung}\ \bibnamefont
		{Kim}}, \bibinfo {author} {\bibfnamefont {Matthew}\ \bibnamefont {Brahlek}},
	\bibinfo {author} {\bibfnamefont {Eliav}\ \bibnamefont {Edrey}}, \ and\
	\bibinfo {author} {\bibfnamefont {Seongshik}\ \bibnamefont {Oh}},\ }\bibfield
{title} {\enquote {\bibinfo {title} {Thickness-independent transport
			channels in topological insulator {{Bi}$_2${Se}$_3$} thin films},}\ }\href
{\doibase 10.1103/PhysRevLett.109.116804} {\bibfield  {journal} {\bibinfo
		{journal} {Phys. Rev. Lett.}\ }\textbf {\bibinfo {volume} {109}},\ \bibinfo
	{pages} {116804} (\bibinfo {year} {2012})}\BibitemShut {NoStop}%
\bibitem [{\citenamefont {J.~W.}\ \emph {et~al.}(2012)\citenamefont {J.~W.},
	\citenamefont {Hsieh}, \citenamefont {Steinberg}, \citenamefont
	{Jarillo-Herrero},\ and\ \citenamefont {Gedik}}]{McIver2012}%
\BibitemOpen
\bibfield  {author} {\bibinfo {author} {\bibfnamefont {McIver}\ \bibnamefont
		{J.~W.}}, \bibinfo {author} {\bibfnamefont {D.}~\bibnamefont {Hsieh}},
	\bibinfo {author} {\bibfnamefont {H.}~\bibnamefont {Steinberg}}, \bibinfo
	{author} {\bibfnamefont {P.}~\bibnamefont {Jarillo-Herrero}}, \ and\ \bibinfo
	{author} {\bibfnamefont {N.}~\bibnamefont {Gedik}},\ }\bibfield  {title}
{\enquote {\bibinfo {title} {Control over topological insulator photocurrents
			with light polarization},}\ }\href {\doibase 10.1038/nnano.2011.214}
{\bibfield  {journal} {\bibinfo  {journal} {Nat. Nanotechnol.}\ }\textbf
	{\bibinfo {volume} {7}},\ \bibinfo {pages} {96--100} (\bibinfo {year}
	{2012})}\BibitemShut {NoStop}%
\bibitem [{\citenamefont {Terzibaschian}\ and\ \citenamefont
	{Enderlein}(1986)}]{Terzibaschian1986}%
\BibitemOpen
\bibfield  {author} {\bibinfo {author} {\bibfnamefont {T.}~\bibnamefont
		{Terzibaschian}}\ and\ \bibinfo {author} {\bibfnamefont {B.}~\bibnamefont
		{Enderlein}},\ }\bibfield  {title} {\enquote {\bibinfo {title} {The
			irreducible representations of the two-dimensional space groups of crystal
			surfaces. theory and applications},}\ }\href {\doibase
	10.1002/pssb.2221330202} {\bibfield  {journal} {\bibinfo  {journal} {physica
			status solidi (b)}\ }\textbf {\bibinfo {volume} {133}},\ \bibinfo {pages}
	{443--461} (\bibinfo {year} {1986})}\BibitemShut {NoStop}%
\bibitem [{\citenamefont {Li}\ \emph {et~al.}(2013)\citenamefont {Li},
	\citenamefont {Tu},\ and\ \citenamefont {Birman}}]{Li2013Mar}%
\BibitemOpen
\bibfield  {author} {\bibinfo {author} {\bibfnamefont {Jian}\ \bibnamefont
		{Li}}, \bibinfo {author} {\bibfnamefont {Jiufeng~J.}\ \bibnamefont {Tu}}, \
	and\ \bibinfo {author} {\bibfnamefont {Joseph~L.}\ \bibnamefont {Birman}},\
}\bibfield  {title} {\enquote {\bibinfo {title} {Symmetry predicted
		transitions in {3D} topological insulators},}\ }\href {\doibase
http://dx.doi.org/10.1016/j.ssc.2013.03.010} {\bibfield  {journal} {\bibinfo
	{journal} {Solid State Communications}\ }\textbf {\bibinfo {volume} {163}},\
\bibinfo {pages} {11 -- 14} (\bibinfo {year} {2013})}\BibitemShut {NoStop}%
\bibitem [{\citenamefont {Slager}\ \emph {et~al.}(2013)\citenamefont {Slager},
	\citenamefont {Mesaros}, \citenamefont {Juricic},\ and\ \citenamefont
	{Zaanen}}]{Slager2013}%
\BibitemOpen
\bibfield  {author} {\bibinfo {author} {\bibfnamefont {Robert-Jan}\
		\bibnamefont {Slager}}, \bibinfo {author} {\bibfnamefont {Andrej}\
		\bibnamefont {Mesaros}}, \bibinfo {author} {\bibfnamefont {Vladimir}\
		\bibnamefont {Juricic}}, \ and\ \bibinfo {author} {\bibfnamefont {Jan}\
		\bibnamefont {Zaanen}},\ }\bibfield  {title} {\enquote {\bibinfo {title} {The
			space group classification of topological band-insulators},}\ }\href
{\doibase 10.1038/nphys2513} {\bibfield  {journal} {\bibinfo  {journal}
		{Nature Phys.}\ }\textbf {\bibinfo {volume} {9}},\ \bibinfo {pages} {98--102}
	(\bibinfo {year} {2013})}\BibitemShut {NoStop}%
\bibitem [{\citenamefont {Lewandowska}\ \emph {et~al.}(2001)\citenamefont
	{Lewandowska}, \citenamefont {Bacewicz}, \citenamefont {Filipowicz},\ and\
	\citenamefont {Paszkowicz}}]{Lewandowska2001}%
\BibitemOpen
\bibfield  {author} {\bibinfo {author} {\bibfnamefont {R.}~\bibnamefont
		{Lewandowska}}, \bibinfo {author} {\bibfnamefont {R.}~\bibnamefont
		{Bacewicz}}, \bibinfo {author} {\bibfnamefont {J.}~\bibnamefont
		{Filipowicz}}, \ and\ \bibinfo {author} {\bibfnamefont {W.}~\bibnamefont
		{Paszkowicz}},\ }\bibfield  {title} {\enquote {\bibinfo {title} {{Raman}
			scattering in {$\alpha$-In$_2$Se$_3$} crystals},}\ }\href {\doibase
	10.1016/S0025-5408(01)00746-2} {\bibfield  {journal} {\bibinfo  {journal}
		{Materials Research Bulletin}\ }\textbf {\bibinfo {volume} {36}},\ \bibinfo
	{pages} {2577 -- 2583} (\bibinfo {year} {2001})}\BibitemShut {NoStop}%
\bibitem [{\citenamefont {Ovander}(1960)}]{Ovander1960}%
\BibitemOpen
\bibfield  {author} {\bibinfo {author} {\bibfnamefont {L.~N.}\ \bibnamefont
		{Ovander}},\ }\bibfield  {title} {\enquote {\bibinfo {title} {The form of the
			{Raman} tensor},}\ }\href@noop {} {\bibfield  {journal} {\bibinfo  {journal}
		{Opt. Spectrosc.}\ }\textbf {\bibinfo {volume} {9}},\ \bibinfo {pages} {302}
	(\bibinfo {year} {1960})}\BibitemShut {NoStop}%
\bibitem [{\citenamefont {Cardona}(1982)}]{Cardona1982}%
\BibitemOpen
\bibfield  {author} {\bibinfo {author} {\bibfnamefont {Manuel}\ \bibnamefont
		{Cardona}},\ }\bibfield  {title} {\enquote {\bibinfo {title} {Resonance
			phenomena},}\ }in\ \href@noop {} {\emph {\bibinfo {booktitle} {Light
			scattering in solids II}}},\ \bibinfo {editor} {edited by\ \bibinfo {editor}
	{\bibfnamefont {Manuel}\ \bibnamefont {Cardona}}\ and\ \bibinfo {editor}
	{\bibfnamefont {Gernot}\ \bibnamefont {G\"{u}ntherodt}}}\ (\bibinfo
{publisher} {Springer-Verlag, Berlin},\ \bibinfo {year} {1982})\ pp.\
\bibinfo {pages} {45--49}\BibitemShut {NoStop}%
\bibitem [{\citenamefont {Koster}(1963)}]{Koster1963}%
\BibitemOpen
\bibfield  {author} {\bibinfo {author} {\bibfnamefont {G.F.}\ \bibnamefont
		{Koster}},\ }\href@noop {} {\emph {\bibinfo {title} {Properties of the
			thirty-two point groups}}},\ Massachusetts institute of technology press
research monograph\ (\bibinfo  {publisher} {M.I.T. Press},\ \bibinfo {year}
{1963})\BibitemShut {NoStop}%
\bibitem [{\citenamefont {Wang}\ and\ \citenamefont {Zhang}(2012)}]{Wang2012}%
\BibitemOpen
\bibfield  {author} {\bibinfo {author} {\bibfnamefont {Bao-Tian}\
		\bibnamefont {Wang}}\ and\ \bibinfo {author} {\bibfnamefont {Ping}\
		\bibnamefont {Zhang}},\ }\bibfield  {title} {\enquote {\bibinfo {title}
		{Phonon spectrum and bonding properties of {Bi$_2$Se$_3$}: Role of strong
			spin-orbit interaction},}\ }\href {http://dx.doi.org/10.1063/1.3689759}
{\bibfield  {journal} {\bibinfo  {journal} {Applied Physics Letters}\
	}\textbf {\bibinfo {volume} {100}},\ \bibinfo {eid} {082109} (\bibinfo {year}
	{2012})}\BibitemShut {NoStop}%
\bibitem [{\citenamefont {Glinka}\ \emph {et~al.}(2015)\citenamefont {Glinka},
	\citenamefont {Babakiray}, \citenamefont {Johnson},\ and\ \citenamefont
	{Lederman}}]{Glinka2015A}%
\BibitemOpen
\bibfield  {author} {\bibinfo {author} {\bibfnamefont {Yuri~D}\ \bibnamefont
		{Glinka}}, \bibinfo {author} {\bibfnamefont {Sercan}\ \bibnamefont
		{Babakiray}}, \bibinfo {author} {\bibfnamefont {Trent~A}\ \bibnamefont
		{Johnson}}, \ and\ \bibinfo {author} {\bibfnamefont {David}\ \bibnamefont
		{Lederman}},\ }\bibfield  {title} {\enquote {\bibinfo {title} {Thickness
			tunable quantum interference between surface phonon and {D}irac plasmon
			states in thin films of the topological insulator {Bi$_2$Se$_3$}},}\ }\href
{http://stacks.iop.org/0953-8984/27/i=5/a=052203} {\bibfield  {journal}
	{\bibinfo  {journal} {Journal of Physics: Condensed Matter}\ }\textbf
	{\bibinfo {volume} {27}},\ \bibinfo {pages} {052203} (\bibinfo {year}
	{2015})}\BibitemShut {NoStop}%
\bibitem [{\citenamefont {Wu}\ \emph {et~al.}(2013)\citenamefont {Wu},
	\citenamefont {Brahlek}, \citenamefont {Valdes~Aguilar}, \citenamefont
	{Stier}, \citenamefont {Morris}, \citenamefont {Lubashevsky}, \citenamefont
	{Bilbro}, \citenamefont {Bansal}, \citenamefont {Oh},\ and\ \citenamefont
	{Armitage}}]{Wu2013}%
\BibitemOpen
\bibfield  {author} {\bibinfo {author} {\bibfnamefont {Liang}\ \bibnamefont
		{Wu}}, \bibinfo {author} {\bibfnamefont {M.}~\bibnamefont {Brahlek}},
	\bibinfo {author} {\bibfnamefont {R.}~\bibnamefont {Valdes~Aguilar}},
	\bibinfo {author} {\bibfnamefont {A.~V.}\ \bibnamefont {Stier}}, \bibinfo
	{author} {\bibfnamefont {C.~M.}\ \bibnamefont {Morris}}, \bibinfo {author}
	{\bibfnamefont {Y.}~\bibnamefont {Lubashevsky}}, \bibinfo {author}
	{\bibfnamefont {L.~S.}\ \bibnamefont {Bilbro}}, \bibinfo {author}
	{\bibfnamefont {N.}~\bibnamefont {Bansal}}, \bibinfo {author} {\bibfnamefont
		{S.}~\bibnamefont {Oh}}, \ and\ \bibinfo {author} {\bibfnamefont {N.~P.}\
		\bibnamefont {Armitage}},\ }\bibfield  {title} {\enquote {\bibinfo {title} {A
			sudden collapse in the transport lifetime across the topological phase
			transition in ({B}i$_{1-x}${I}n$_x$)$_2${Se}$_3$},}\ }\href
{http://dx.doi.org/10.1038/nphys2647} {\bibfield  {journal} {\bibinfo
		{journal} {Nature Phys.}\ }\textbf {\bibinfo {volume} {9}},\ \bibinfo {pages}
	{410--414} (\bibinfo {year} {2013})}\BibitemShut {NoStop}%
\bibitem [{\citenamefont {Lee}\ \emph {et~al.}(2014)\citenamefont {Lee},
	\citenamefont {Xu}, \citenamefont {Shubeita}, \citenamefont {Brahlek},
	\citenamefont {Koirala}, \citenamefont {Oh},\ and\ \citenamefont
	{Gustafsson}}]{Lee2014}%
\BibitemOpen
\bibfield  {author} {\bibinfo {author} {\bibfnamefont {Hang~Dong}\
		\bibnamefont {Lee}}, \bibinfo {author} {\bibfnamefont {Can}\ \bibnamefont
		{Xu}}, \bibinfo {author} {\bibfnamefont {Samir~M.}\ \bibnamefont {Shubeita}},
	\bibinfo {author} {\bibfnamefont {Matthew}\ \bibnamefont {Brahlek}}, \bibinfo
	{author} {\bibfnamefont {Nikesh}\ \bibnamefont {Koirala}}, \bibinfo {author}
	{\bibfnamefont {Seongshik}\ \bibnamefont {Oh}}, \ and\ \bibinfo {author}
	{\bibfnamefont {Torgny}\ \bibnamefont {Gustafsson}},\ }\bibfield  {title}
{\enquote {\bibinfo {title} {Indium and bismuth interdiffusion and its
			influence on the mobility in {In$_2$Se$_3$}/{Bi$_2$Se$_3$}},}\ }\href
{\doibase http://dx.doi.org/10.1016/j.tsf.2014.01.082} {\bibfield  {journal}
	{\bibinfo  {journal} {Thin Solid Films}\ }\textbf {\bibinfo {volume} {556}},\
	\bibinfo {pages} {322 -- 324} (\bibinfo {year} {2014})}\BibitemShut {NoStop}%
\bibitem [{\citenamefont {{Shastry}}\ and\ \citenamefont
	{{Shraiman}}(1991)}]{Shastry1991}%
\BibitemOpen
\bibfield  {author} {\bibinfo {author} {\bibfnamefont {B.~S.}\ \bibnamefont
		{{Shastry}}}\ and\ \bibinfo {author} {\bibfnamefont {B.~I.}\ \bibnamefont
		{{Shraiman}}},\ }\bibfield  {title} {\enquote {\bibinfo {title} {{Raman
				Scattering in {Mott-Hubbard} Systems}},}\ }\href {\doibase
	10.1142/S0217979291000237} {\bibfield  {journal} {\bibinfo  {journal} {Int.
			J. Mod. Phys. B}\ }\textbf {\bibinfo {volume} {5}},\ \bibinfo {pages}
	{365--388} (\bibinfo {year} {1991})}\BibitemShut {NoStop}%
\bibitem [{\citenamefont {Khveshchenko}\ and\ \citenamefont
	{Wiegmann}(1994)}]{Khveshchenko1994}%
\BibitemOpen
\bibfield  {author} {\bibinfo {author} {\bibfnamefont {D.~V.}\ \bibnamefont
		{Khveshchenko}}\ and\ \bibinfo {author} {\bibfnamefont {P.~B.}\ \bibnamefont
		{Wiegmann}},\ }\bibfield  {title} {\enquote {\bibinfo {title} {Raman
			scattering and anomalous current algebra in mott insulators},}\ }\href
{\doibase 10.1103/PhysRevLett.73.500} {\bibfield  {journal} {\bibinfo
		{journal} {Phys. Rev. Lett.}\ }\textbf {\bibinfo {volume} {73}},\ \bibinfo
	{pages} {500--503} (\bibinfo {year} {1994})}\BibitemShut {NoStop}%
\bibitem [{\citenamefont {Cheng}\ and\ \citenamefont {Ren}(2011)}]{Chen2011}%
\BibitemOpen
\bibfield  {author} {\bibinfo {author} {\bibfnamefont {Wei}\ \bibnamefont
		{Cheng}}\ and\ \bibinfo {author} {\bibfnamefont {Shang-Fen}\ \bibnamefont
		{Ren}},\ }\bibfield  {title} {\enquote {\bibinfo {title} {Phonons of single
			quintuple {Bi${}_{2}$Te${}_{3}$} and {Bi${}_{2}$Se${}_{3}$} films and bulk
			materials},}\ }\href {\doibase 10.1103/PhysRevB.83.094301} {\bibfield
	{journal} {\bibinfo  {journal} {Phys. Rev. B}\ }\textbf {\bibinfo {volume}
		{83}},\ \bibinfo {pages} {094301} (\bibinfo {year} {2011})}\BibitemShut
{NoStop}%
\bibitem [{\citenamefont {Bianchi}\ \emph {et~al.}(2012)\citenamefont
	{Bianchi}, \citenamefont {Hatch}, \citenamefont {Guan}, \citenamefont
	{Planke}, \citenamefont {Mi}, \citenamefont {Iversen},\ and\ \citenamefont
	{Hofmann}}]{Bianchi2012}%
\BibitemOpen
\bibfield  {author} {\bibinfo {author} {\bibfnamefont {Marco}\ \bibnamefont
		{Bianchi}}, \bibinfo {author} {\bibfnamefont {Richard~C}\ \bibnamefont
		{Hatch}}, \bibinfo {author} {\bibfnamefont {Dandan}\ \bibnamefont {Guan}},
	\bibinfo {author} {\bibfnamefont {Tilo}\ \bibnamefont {Planke}}, \bibinfo
	{author} {\bibfnamefont {Jianli}\ \bibnamefont {Mi}}, \bibinfo {author}
	{\bibfnamefont {Bo~Brummerstedt}\ \bibnamefont {Iversen}}, \ and\ \bibinfo
	{author} {\bibfnamefont {Philip}\ \bibnamefont {Hofmann}},\ }\bibfield
{title} {\enquote {\bibinfo {title} {The electronic structure of clean and
			adsorbate-covered {Bi$_{2}$Se$_{3}$} : an angle-resolved photoemission
			study},}\ }\href {http://stacks.iop.org/0268-1242/27/i=12/a=124001}
{\bibfield  {journal} {\bibinfo  {journal} {Semiconductor Science and
			Technology}\ }\textbf {\bibinfo {volume} {27}},\ \bibinfo {pages} {124001}
	(\bibinfo {year} {2012})}\BibitemShut {NoStop}%
\bibitem [{\citenamefont {Menshchikova}\ \emph {et~al.}(2011)\citenamefont
	{Menshchikova}, \citenamefont {Eremeev},\ and\ \citenamefont
	{Chulkov}}]{Menshchikova2011}%
\BibitemOpen
\bibfield  {author} {\bibinfo {author} {\bibfnamefont {T.~V.}\ \bibnamefont
		{Menshchikova}}, \bibinfo {author} {\bibfnamefont {S.~V.}\ \bibnamefont
		{Eremeev}}, \ and\ \bibinfo {author} {\bibfnamefont {E.~V.}\ \bibnamefont
		{Chulkov}},\ }\bibfield  {title} {\enquote {\bibinfo {title} {On the origin
			of two-dimensional electron gas states at the surface of topological
			insulators},}\ }\href {http://dx.doi.org/10.1134/S0021364011140104}
{\bibfield  {journal} {\bibinfo  {journal} {JETP Letters}\ }\textbf {\bibinfo
		{volume} {94}} (\bibinfo {year} {2011})}\BibitemShut {NoStop}%
\bibitem [{\citenamefont {Fano}(1961)}]{Fano1961}%
\BibitemOpen
\bibfield  {author} {\bibinfo {author} {\bibfnamefont {U.}~\bibnamefont
		{Fano}},\ }\bibfield  {title} {\enquote {\bibinfo {title} {{Effects of
				Configuration Interaction on Intensities and Phase Shifts}},}\ }\href
{\doibase 10.1103/PhysRev.124.1866} {\bibfield  {journal} {\bibinfo
		{journal} {Phys. Rev.}\ }\textbf {\bibinfo {volume} {124}},\ \bibinfo {pages}
	{1866} (\bibinfo {year} {1961})}\BibitemShut {NoStop}%
\bibitem [{\citenamefont {Klein}(1983)}]{Klein1983}%
\BibitemOpen
\bibfield  {author} {\bibinfo {author} {\bibfnamefont {M.V.}\ \bibnamefont
		{Klein}},\ }\bibfield  {title} {\enquote {\bibinfo {title} {Electronic raman
			scattering},}\ }in\ \href@noop {} {\emph {\bibinfo {booktitle} {Light
			Scattering in Solids I}}},\ \bibinfo {editor} {edited by\ \bibinfo {editor}
	{\bibfnamefont {M.}~\bibnamefont {Cardona}}\ and\ \bibinfo {editor}
	{\bibfnamefont {G.}~\bibnamefont {G\"{u}ntherodt}}}\ (\bibinfo  {publisher}
{Springer-Verlag, Berlin},\ \bibinfo {year} {1983})\ pp.\ \bibinfo {pages}
{169--172}\BibitemShut {NoStop}%
\bibitem [{\citenamefont {Sobota}\ \emph {et~al.}(2013)\citenamefont {Sobota},
	\citenamefont {Yang}, \citenamefont {Kemper}, \citenamefont {Lee},
	\citenamefont {Schmitt}, \citenamefont {Li}, \citenamefont {Moore},
	\citenamefont {Analytis}, \citenamefont {Fisher}, \citenamefont {Kirchmann},
	\citenamefont {Devereaux},\ and\ \citenamefont {Shen}}]{Sobota2013}%
\BibitemOpen
\bibfield  {author} {\bibinfo {author} {\bibfnamefont {J.~A.}\ \bibnamefont
		{Sobota}}, \bibinfo {author} {\bibfnamefont {S.-L.}\ \bibnamefont {Yang}},
	\bibinfo {author} {\bibfnamefont {A.~F.}\ \bibnamefont {Kemper}}, \bibinfo
	{author} {\bibfnamefont {J.~J.}\ \bibnamefont {Lee}}, \bibinfo {author}
	{\bibfnamefont {F.~T.}\ \bibnamefont {Schmitt}}, \bibinfo {author}
	{\bibfnamefont {W.}~\bibnamefont {Li}}, \bibinfo {author} {\bibfnamefont
		{R.~G.}\ \bibnamefont {Moore}}, \bibinfo {author} {\bibfnamefont {J.~G.}\
		\bibnamefont {Analytis}}, \bibinfo {author} {\bibfnamefont {I.~R.}\
		\bibnamefont {Fisher}}, \bibinfo {author} {\bibfnamefont {P.~S.}\
		\bibnamefont {Kirchmann}}, \bibinfo {author} {\bibfnamefont {T.~P.}\
		\bibnamefont {Devereaux}}, \ and\ \bibinfo {author} {\bibfnamefont {Z.-X.}\
		\bibnamefont {Shen}},\ }\bibfield  {title} {\enquote {\bibinfo {title}
		{Direct optical coupling to an unoccupied {D}irac surface state in the
			topological insulator {Bi}$_2${Se}$_3$},}\ }\href {\doibase
	10.1103/PhysRevLett.111.136802} {\bibfield  {journal} {\bibinfo  {journal}
		{Phys. Rev. Lett.}\ }\textbf {\bibinfo {volume} {111}},\ \bibinfo {pages}
	{136802} (\bibinfo {year} {2013})}\BibitemShut {NoStop}%
\bibitem [{\citenamefont {Niesner}\ \emph {et~al.}(2012)\citenamefont
	{Niesner}, \citenamefont {Fauster}, \citenamefont {Eremeev}, \citenamefont
	{Menshchikova}, \citenamefont {Koroteev}, \citenamefont {Protogenov},
	\citenamefont {Chulkov}, \citenamefont {Tereshchenko}, \citenamefont {Kokh},
	\citenamefont {Alekperov}, \citenamefont {Nadjafov},\ and\ \citenamefont
	{Mamedov}}]{Niesner2012}%
\BibitemOpen
\bibfield  {author} {\bibinfo {author} {\bibfnamefont {D.}~\bibnamefont
		{Niesner}}, \bibinfo {author} {\bibfnamefont {Th.}\ \bibnamefont {Fauster}},
	\bibinfo {author} {\bibfnamefont {S.~V.}\ \bibnamefont {Eremeev}}, \bibinfo
	{author} {\bibfnamefont {T.~V.}\ \bibnamefont {Menshchikova}}, \bibinfo
	{author} {\bibfnamefont {Yu.~M.}\ \bibnamefont {Koroteev}}, \bibinfo {author}
	{\bibfnamefont {A.~P.}\ \bibnamefont {Protogenov}}, \bibinfo {author}
	{\bibfnamefont {E.~V.}\ \bibnamefont {Chulkov}}, \bibinfo {author}
	{\bibfnamefont {O.~E.}\ \bibnamefont {Tereshchenko}}, \bibinfo {author}
	{\bibfnamefont {K.~A.}\ \bibnamefont {Kokh}}, \bibinfo {author}
	{\bibfnamefont {O.}~\bibnamefont {Alekperov}}, \bibinfo {author}
	{\bibfnamefont {A.}~\bibnamefont {Nadjafov}}, \ and\ \bibinfo {author}
	{\bibfnamefont {N.}~\bibnamefont {Mamedov}},\ }\bibfield  {title} {\enquote
	{\bibinfo {title} {Unoccupied topological states on bismuth chalcogenides},}\
}\href {\doibase 10.1103/PhysRevB.86.205403} {\bibfield  {journal} {\bibinfo
	{journal} {Phys. Rev. B}\ }\textbf {\bibinfo {volume} {86}},\ \bibinfo
{pages} {205403} (\bibinfo {year} {2012})}\BibitemShut {NoStop}%
\end{thebibliography}

%
\end{document}